\documentclass[floatfix,a4paper,10pt,aps,pre,longbibliography,twocolumn,amsmath,amssymb,nofootinbib,superscriptaddress]{revtex4-2}
\usepackage{xcolor,hyperref}
\hypersetup{
   colorlinks,
   linkcolor={blue!50!black},%{red!80!black},
   citecolor={blue!50!black},
   urlcolor={blue!80!black}
}
\usepackage{epsfig, amsmath}
\usepackage{bm}
\usepackage{soul}
\usepackage{hyperref}
\usepackage{footmisc}
\DeclareMathAlphabet{\mathitbf}{OML}{cmm}{b}{it}

\newcommand{\rv}{\mathitbf r}

\begin{document}

\title{Mechanical excitation and marginal triggering during avalanches in sheared amorphous solids}
\author{D.~Richard}
\thanks{contributed equally}
\affiliation{Institute for Theoretical Physics, University of Amsterdam, Science Park 904, Amsterdam, Netherlands}
\affiliation{Department of Physics and BioInspired Institute, Syracuse University, Syracuse, NY 13244}
\affiliation{Univ. Grenoble Alpes, CNRS, LIPhy, 38000 Grenoble, France}
\author{A.~Elgailani}
\thanks{contributed equally}
\affiliation{Northeastern University, Boston, Massachusetts 02115, USA}
\author{D.~Vandembroucq}
\affiliation{PMMH, CNRS UMR 7636, ESPCI Paris, PSL University, Sorbonne  Université, Université  de Paris, F-75005 Paris, France}
\author{M.L.~Manning}
\affiliation{Department of Physics and BioInspired Institute, Syracuse University, Syracuse, NY 13244}
\author{C.E. ~Maloney}
\affiliation{Northeastern University, Boston, Massachusetts 02115, USA}

\begin{abstract}
We study plastic strain during individual avalanches in overdamped particle-scale molecular dynamics (MD) and meso-scale elasto-plastic models (EPM) for amorphous solids sheared in the athermal quasi-static limit.  We show that the spatial correlations in plastic activity exhibit a short lengthscale that grows as $t^{3/4}$ in MD and ballistically in EPM, and is generated by mechanical excitation of nearby sites not necessarily close to their stability thresholds, and a longer lengthscale that grows diffusively for both models and is associated with remote marginally stable sites. These similarities in spatial correlations explain why simple EPMs accurately capture the size distribution of avalanches observed in MD, though the temporal profiles and dynamical critical exponents are quite different.
\end{abstract}

\maketitle

Many driven disordered solids, ranging from glasses to granular matter to magnetic systems, respond via complex avalanches that are difficult to predict \cite{durin2000scaling,dalton2001self,maaloy2006local}. A better understanding of these dynamics, even in simple model systems, would aid in avalanche detection and material design.

We focus here on amorphous solids subject to slowly imposed shear, which fail via a broad spectrum of avalanches of plastic deformation caused by redistribution of stress after local yielding events~\cite{sethna2001crackling,sun2010plasticity,lin2015criticality,Budrikis-NatComm17}. The highly anisotropic and long-ranged nature of the stress redistribution leads to a characteristic structure of the avalanches, where correlations are strongest along the directions of maximum imposed shear. Most previous work has focused on the distribution of avalanche sizes~\cite{Talamali-PRE11,Salerno-PRL12,lin2014scaling,liu2016driving,Budrikis-NatComm17,clemmer2021criticalityp1,clemmer2021criticalityp2} and the spatial correlations of plastic strain which develop over the course of successive avalanches~\cite{maloney2009anisotropic,TALAMALI2012275,chattoraj2013elastic,nicolas2014spatiotemporal,puosi2016plastic}.
More recently, the dissipation rate as a function of time during individual avalanches was studied in experiments on bulk metallic glass (BMG) pillars~\cite{Dahmen-PRL14}, and in a computer model~\cite{liu2016driving}. The temporal response was observed to be similar to that in previously explored dynamically critical systems~\cite{sethna2001crackling}, and explained using a mean-field theory~\cite{Dahmen-PRL14}. However, surprisingly, no work, in either experiment or simulation, has yet characterized how the individual avalanches proceed in time \emph{and space}. 

To address this question we turn to computational models, including particle-based simulations, such as molecular dynamics (MD) and related energy minimization techniques, which have been the workhorse for modeling sheared amorphous solids for decades, as well as elasto-plastic models (EPMs). EPMs assume an amorphous solid is composed of meso-scale regions that will yield when the local stress reaches a specified threshold. There are many different versions of EPMs~\cite{nicolas2018deformation} that differ in how they introduce disorder, evolve propogating stress fields, etc. In perhaps the simplest class of EPM~\cite{baret-PRL02,Talamali-PRE11,Budrikis-PRE13,Budrikis-NatComm17,Tyukodi-PRL18} the system is evolved quasi-statically, and, after any instability, the stresses are fully equilibrated over all space -- effectively instantaneously -- before allowing for any subsequent yielding. This is in contrast to MD simulations where, after a local rearrangement, the stress change propagates continuously in time and space -- diffusively for overdamped systems~\cite{puosi2014time} and ballistically for underdamped systems -- away from the plastic instability. 

Despite the fact that the quasi-static EPMs are completely devoid of any realistic description of the dynamics of stress redistribution, they capture the critical scaling exponents observed in overdamped MD simulations~\cite{Salerno-PRL12,Salerno-PRE13,Tyukodi-PRE19}. They are therefore, in some sense, unreasonably good, and the reason for their fidelity demands an explanation.

In this letter, we compare two systems: an overdamped, gradient-descent MD simulation and a quasi-static EPM.
We show that while the temporal profile of the avalanches is quite different in the two models -- the EPM agrees with previously known mean field results while the MD does not -- the spatial structure of the correlations that develop is strikingly similar. The spatial correlation function can be characterized by two distinct lengthscales: 
a short length, $\xi_{\text{excite}}$, and a longer one, $\xi_{\text{marginal}}$.
$\xi_{\text{excite}}$ grows ballistically in time for the EPM, and we argue that the reason for this ballistic propagation is the ``mechanical excitation" mechanism suggested by Idema and Liu~\cite{idema2014mechanical}: one event generates a stress redistribution that causes nearby sites to exceed their threshold for stability, triggering new events. If the timescale over which an unstable site transforms is short compared to the propagation of the stress, then the resulting dynamics are reminiscent of toppling dominos, with a ballistic wavespeed equal to the spacing of the dominos divided by the ``toppling time" -- the time it takes one, once destabilized, to fall onto its neighbor. $\xi_{\text{excite}}$ grows less quickly in time for the MD, but it is still strongly super-diffusive, suggesting that a similar mechanism is at play and hinting that the ``toppling time" in MD is more complicated than in EPM.
In contrast, $\xi_{\text{marginal}}$ grows diffusively in time for both models, and it has a pronounced system-size dependence. At long distance the low positive and negative contributions of the quadrupolar stress interactions add up as a mechanical noise and are expected to drive local zones close to marginality~\cite{Sollich-PRL97,Lemaitre-Caroli-PRE07,Agoritsas-EPJE15,Wyart-PRX16,ferrero2019criticality}. The observed size dependence suggests that $\xi_{\text{marginal}}$ is the length scale at which the weakest sites in the system are close enough to the triggering event to be destabilized.

Our MD glass former consists of a generic two dimensional binary mixture~\cite{lerner2019mechanical}. Samples are deformed using standard athermal quasistatic shear protocol~\cite{maloney2006amorphous}. At the onset of each instability, we trigger the avalanche by affinely deforming the system and subsequently let the system relax via gradient descent dynamics. More details are given in the SM. In order to break up large avalanche into a sequence of individual plastic events, we employ a recently proposed persistent homology method~\cite{stanifer2021avalanche}, detailed in the SM~\cite{SM}.

For the EPM, we use the same shear transformation based model in \cite{Khirallah2021yielding} with the same initialization and evolution rules but with different loading -- forward shearing instead of cyclic, described in more detail in the SM~\cite{SM}. The dynamical update rules under an applied global shear strain are (i) for a given stress field, synchronously allow all sites over threshold to yield and recompute the stress field everywhere; (ii) repeat (i) until all sites are below threshold; and (iii) advance the globally applied total strain until precisely one site is at its stability threshold. The synchronous update of unstable sites defined in step (i) sets the time unit of the model.

%%%%%%%%%%%%%%%%%%%%%%%%%%%%%%%% Figure 1 %%%%%%%%%%%%%%%%%%%%%%%%%%%%%
\begin{figure}[t!]
  \includegraphics[width = 1.\columnwidth]{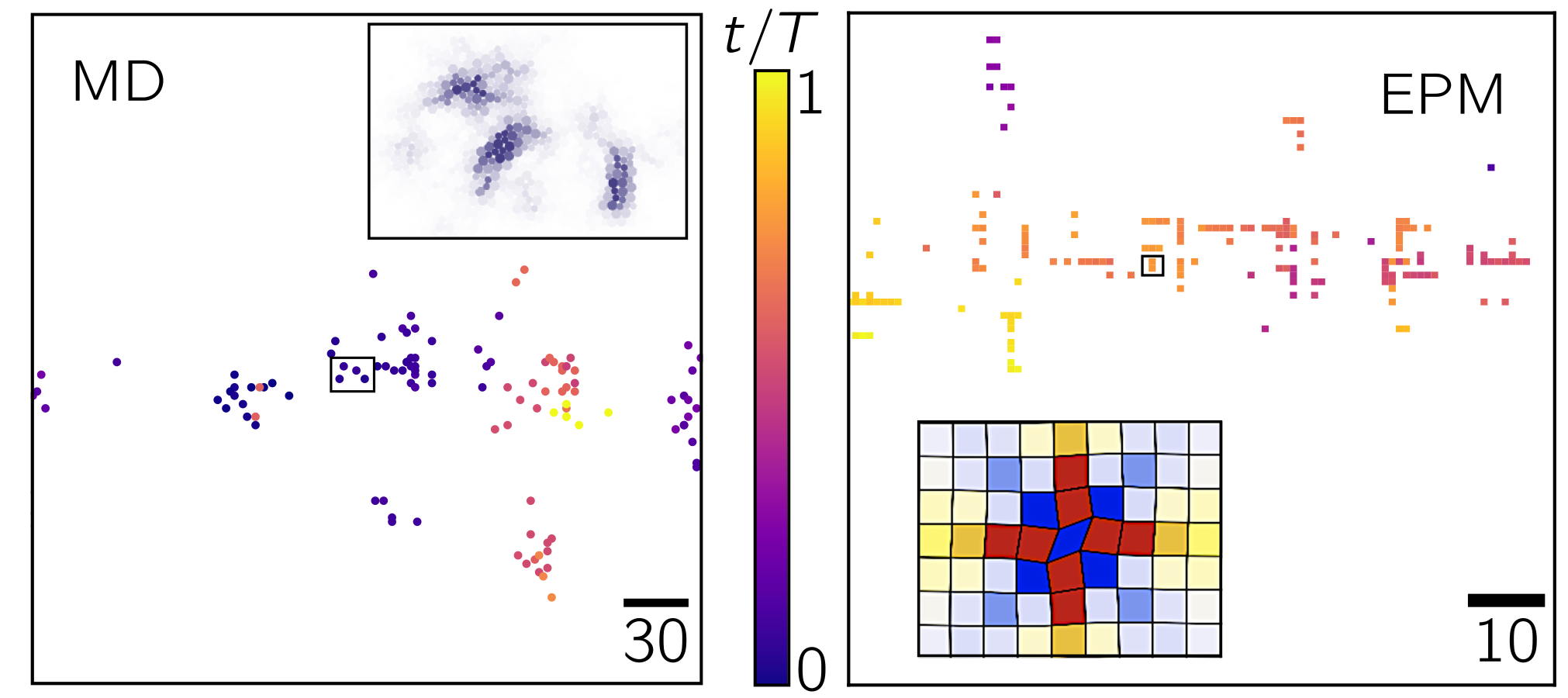} %scale=1.0
  \caption{\textbf{Spatio-temporal plastic evolution.} Typical spatio-temporal map of an avalanche in Molecular Dynamics (MD) with gradient descent dynamics (left) and in the Elasto-plastic Model (EPM) with synchronous dynamics (right). Each plastic event is colored according to its normalized birth time $t/T$, with $T$ the avalanche duration. Insets show the particle resolved plastic field (left) and EPM local stress redistribution (right)}
  \label{fig:intro}
\end{figure}
%%%%%%%%%%%%%%%%%%%%%%%%%%%%%%%%%%%%%%%%%%%%%%%%%%%%%%%%%%%%%%%%%%%%%%

We first qualitatively describe a typical avalanche in both models. Fig.~\ref{fig:intro} highlights individual sites that have yielded, colored according to the time at which the site yielded. The pattern which emerges is a set of clusters, where all sites within a cluster are nearly the same color, extend along a direction of maximum shear, and almost continuously fill space.

In contrast, these larger clusters are typically separated from one another by gaps of material. Nearby clusters are not necessarily triggered sequentially in time, suggesting that gaps between clusters correspond to regions of material that are stable enough to survive the increased local stresses at the edges of the growing cluster. The same intermittent activity is found in depinning avalanches with long-ranged interactions~\cite{laurson2010avalanches,le2021spatial}.

%%%%%%%%%%%%%%%%%%%%%%%%%%%%%%%% Figure 2 %%%%%%%%%%%%%%%%%%%%%%%%%%%%%
\begin{figure}[t!]
  \includegraphics[width = 1.\columnwidth]{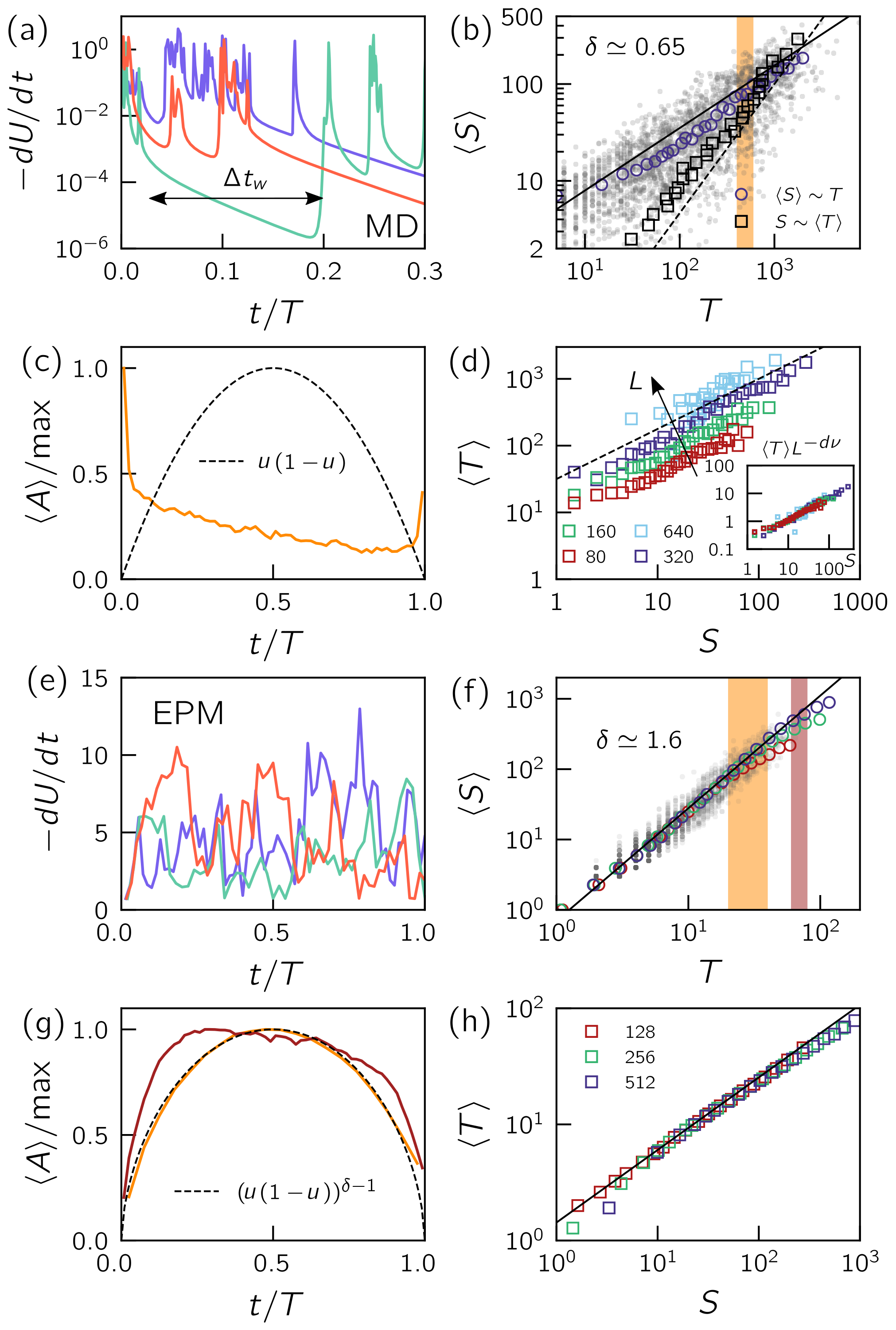} %scale=1.0
    \caption{\textbf{Plastic activity profile.} (a) Typical dissipation rate $-dU/dt$ in MD plotted as a function of the reduced time $t/T$ for duration $1500<T<3000$. (b) Scatter plot of avalanche size versus duration. The Blue and black empty symbols are running means $\langle S \rangle \sim T$ and $\langle T \rangle \sim S$, respectively. Solid and dashed lines indicate the scaling $\langle S \rangle \sim T^{\delta}$, with exponent $\delta=0.65$ and $\delta=4/3$, respectively. (c) Average normalized activity for $400<T<600$ (orange region in (b)). (d) Average duration $\langle T \rangle$ at a fixed avalanche size $S$ for different system size. Insets show the same data collapsed using $\langle T \rangle L^{-d\nu}$, with $\nu\simeq0.4$ and $d$ the spatial dimension. Panels (e), (f), (g) and (h) show the same results as in (a), (b), (c), and (d) but for the EPM where $-dU/dt$ is measured as the energy dissipation per sweep.}
  \label{fig:activity}
\end{figure}
%%%%%%%%%%%%%%%%%%%%%%%%%%%%%%%%%%%%%%%%%%%%%%%%%%%%%%%%%%%%%%%%%%%%%%

Given this qualitative similarity in the spatio-temporal structure, and the utility of mean-field models in predicting temporal dynamics in depinning and other disordered critical phenomena, we next quantitatively analyze the temporal activity profiles in MD and EPM. In Fig.\ref{fig:activity}(a) and (e), we plot the dissipation rate $-dU/dt$ as a function of the normalized duration $t/T$ for a few typical avalanches. In MD, we find that the activity exhibits intermittent bursts of activity followed by long inactive periods, so that the activity magnitude spans six decades. Inactive regions are associated with marginal sites that are only weakly unstable. For a typical saddle node bifurcation, we expect that the waiting time $\Delta t_w$ before activation will scale as $\Delta t_w \sim (\sigma_i - \sigma_i^{\rm th})^{-1/2}$ \cite{strogatz2018nonlinear}, with $\sigma_i$ and $\sigma_i^{\rm th}$ the local stress and local yield stress of the unstable site $i$. In contrast, unstable sites in our EPM transform instantaneously regardless how far they are over threshold, termed a uniform activation rate. As a consequence, activity fluctuations in EPM (Fig.\ref{fig:activity}(e)) vary by less than an order of magnitude.

Another difference between the MD and EPM can be seen by plotting the avalanche size $S$ versus duration $T$, Fig.\ref{fig:activity}(b) for MD and (f) for EPM. We find that MD data scatter much more than our EPM. This is similar to previous results comparing an EPM with a uniform activation rate to one with a so-called progressive rate model, where the activation rate is a function of the overshoot $\sigma_i - \sigma_i^{\rm th}$ \cite{aguirre2018critical,ferrero2019criticality}. In both MD and EPM, the average size for a fixed duration show a power law $\langle S \rangle \propto T^{\delta}$, where the dynamical exponent $\delta=d_f/z$ with $d_f$ and $z$ the static and dynamical fractal exponent, respectively. Independent estimation of $d_f$ and $z$ are reported in the Supplemental Material (SM)~\cite{SM} and consistent with previous works~\cite{clemmer2021criticalityp2,liu2016driving}. The EPM gives a larger $\delta$ than MD as the dynamical exponent $z\simeq0.6$ is lower than MD ($z\simeq1.55$), although it is consistent with previous EPM-type models~\cite{liu2016driving}, which used continuous-in-time dynamics rather than the automaton rules we apply here.

More surprisingly, in MD we find a discrepancy between average avalanche size for a fixed duration $\langle S \rangle_T$ and the average duration for a fixed size $\langle T \rangle_S$. We find that the average duration $\langle T \rangle_S$ is system-size dependent and grows with $L$ as $L^{d\nu}$, see Fig.\ref{fig:activity}(d). We can understand this size effect in terms of the large waiting time $\Delta t_w$ associated with unstable sites that are just barely over threshold, $\sigma_i \gtrsim \sigma_i^{\rm th}$. We expect that the overshoot $\langle \sigma_i - \sigma_i^{\rm th} \rangle$ will decrease with system size as $\sim L^{-\frac{d}{1+\eta}}$. This scaling of the characteristic overshoot for unstable sites is different from the previously reported scaling for thresholds associated with marginal stable sites~\cite{karmakar2010statistical,lin2014density,Tyukodi-PRE19,ruscher2020residual}. The average waiting time caused by weak triggering will follow $\langle \Delta t_w \rangle \sim L^{\frac{d}{2(1+\eta)}}\sim L^{d\nu}$. We find a good collapse of MD data for $\nu\simeq 0.4$, giving $\eta\simeq 1/4$. A similar finite-size effect, though with a different exponent, was previously noticed in progressive rate EPMs~\cite{aguirre2018critical,ferrero2019criticality}. 

In Fig.\ref{fig:activity}(c) and (g), we show the normalized average activity profile $\langle A \rangle/\langle A \rangle_{\rm max}$ (with $A=-dU/dt$) plotted against the reduced time $t/T$ in the MD and EPM respectively. For the EPM and avalanches within the scaling regime, we find that profiles are symmetrical and well modelled by $f(u)=(u(1-u))^{\delta-1}$, with $u=t/T$, consistent with previous works and mean field theories~\cite{laurson2013evolution,liu2016driving}. For large avalanches in EPM, with long duration beyond the scaling regime where system size effects become relevant, one finds a profile which departs from that scaling and is skewed with more activity at early times. A similar transition to skewed profiles for large evens was recently observed in granular flows \cite{baldassarri2019breakdown}. In MD, the profile deviates strongly from the $f(u)=(u(1-u))^{\delta-1}$ form. This is due to inactive periods with almost zero dissipation. The average activity at the mid-point of the avalanche is systematically lower than at its beginning, the latter being by definition always active. We speculate that one would recover a symmetrical activity profile by introducing inertia in the dynamics, which would facilitate barrier crossing in weakly unstable regions.

%%%%%%%%%%%%%%%%%%%%%%%%%%%%%%%% Figure 3 %%%%%%%%%%%%%%%%%%%%%%%%%%%%%
\begin{figure*}[t!]
  \includegraphics{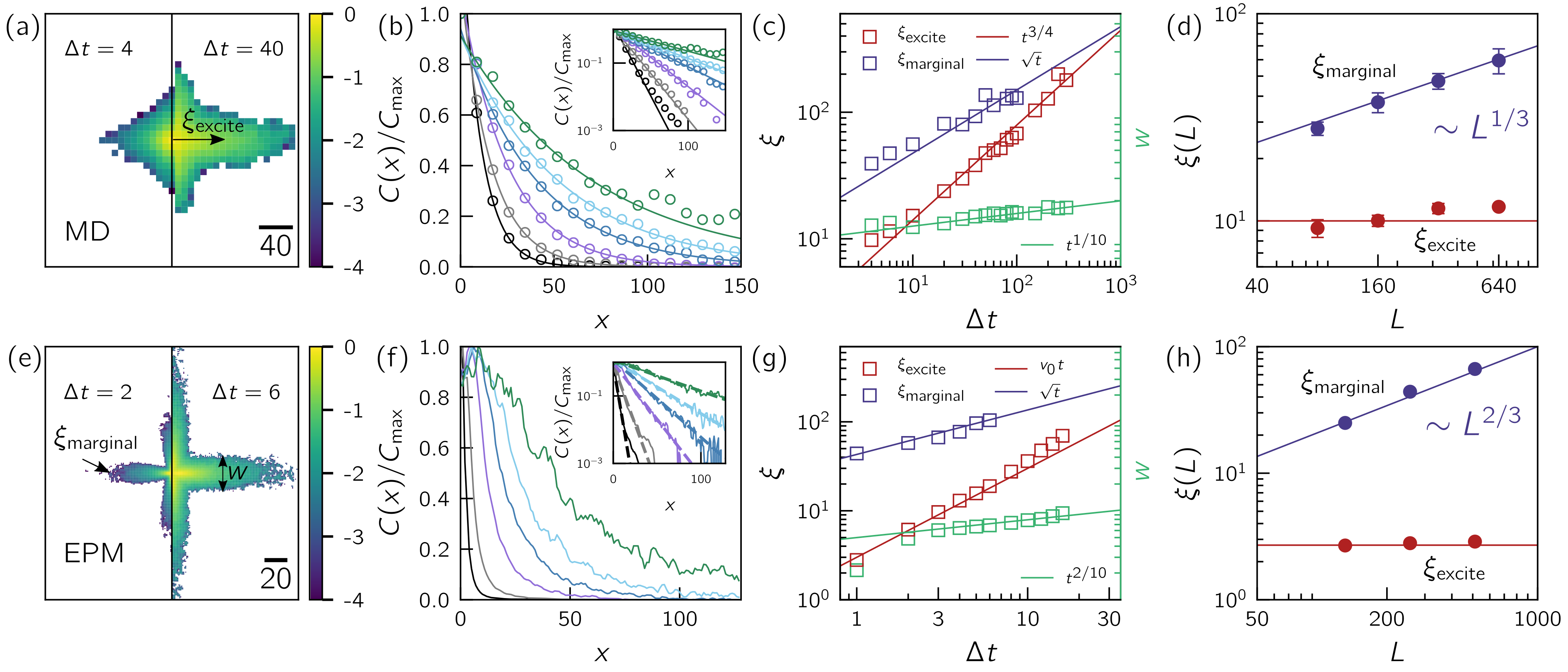} %scale=1.0
    \caption{\textbf{Spatio-temporal plastic propagation.} (a) particle-based (MD) 2-points 2-times normalized correlation function $C(x)/C_{\rm max}$ in $\log_{10}$ for $\Delta t =4$ and $40$. White regions correspond to a negative correlation. $\xi_{\rm excite}$ and $\xi_{\rm marginal}$ correspond to the short length scale decay and negative crossing, respectively. $w$ indicates the width of the plastic localization. (b) Correlation function along the main stress redistribution for different delay time $\Delta t=4$, $10$, $20$, $40$, $60$, and $100$. Inset shows the correlation in semi-log scale. Solid lines are fits of the form $C(x)/C_{\rm max}\sim e^{-x/\xi_{\rm excite}}$. (c) $\xi_{\rm excite}$ (red) and $\xi_{\rm marginal}$ (blue) plotted against $\Delta t$. The blue line corresponds to the diffusive response associated with the stress field generated by a single event at $\Delta t=0$. The red line indicates a superdiffusive regime with $\sim t^{3/4}$. Green data are the plastic width $w$ extracted from fitting the normal correlation $C(y)$ (away from the center) by $C(y)\sim e^{-(y/w)^2/2}$. The green line indicates a subdiffusive regime with $\sim t^{1/10}$. (d) $\xi_{\rm excite}$ and $\xi_{\rm marginal}$ for $\Delta t=t_{\rm STZ}=6$ plotted as a function of the system size $L$. $\xi_{\rm marginal}(L)$ scales as $\sim L^{1/(1+\nu)}$, with $\nu=2$ an effective pseudo-gap exponent. (e), (f), (g), and (h) are the same results as in (a), (b), (c) and (d) but for our mesoscale model (EPM) with $\Delta t=1$, $2$, $4$, $6$, $8$, and $12$. The solid red line in (g) indicates $v_0 t$, with $v_0=3$. Results in (h) are for $\Delta t=1$ where $\xi_{\rm marginal}(L)$ scales as $\sim L^{1/(1+\nu)}$, with $\nu=2/3$. $\xi_{\rm excite}(L)$ is fairly constant for both MD and EPM.}
  \label{fig:plasticlength}
\end{figure*}
%%%%%%%%%%%%%%%%%%%%%%%%%%%%%%%%%%%%%%%%%%%%%%%%%%%%%%%%%%%%%%%%%%%%%%

We next analyze the spatial correlations that build up during the avalanche. To do this, we define a 2-points 2-times correlation function $C(r_0,r_0+\vec{r},t_0,t_0+\Delta t)=\langle \overline{\Delta P(r_0,t_0)\Delta P(r_0+\vec{r},t_0+\Delta t)} \rangle$, with $\Delta P = P-\overline{P}$ and where $\overline{...}$ and $\langle ...\rangle$ represent a spatial and ensemble average, respectively. Here, $P(t)$ corresponds to the incremental plastic field measured at a given time $t$. In order to improve the statistics of the particle-based correlations, we assume time-translational invariance and average over $t_0$. Data supporting this assumption for both MD and EPM is shown in SM Sec.~6~\cite{SM}.

Fig.~\ref{fig:plasticlength}(a) shows the MD normalized correlation $C(x,y)/C_{\rm max}$ for delay times $\Delta t=4$ (left) and $40$ (right). There are positive correlations along the directions of imposed shear (the $x$ and $y$ axes). The spatial extent of the region of positive correlation grows with $\Delta t$ \cite{notesymmetry}. The same data are shown for the EPM in Fig.~\ref{fig:plasticlength}(e) for $\Delta t=2$ and $6$ where we observe the same qualitative behavior. 

In Fig.~\ref{fig:plasticlength} (b) and (f), we plot $C(x)$ along the $x$-axis at various $\Delta t$. $C(x)$ decays exponentially at small $x$, and we use the exponential decay rate to define a short-range lengthscale $\xi_{\text{excite}}$. Despite the initial exponential decay, $C(x)$ crosses through zero at a finite $x$, indicating anti-correlation, and we use this zero crossing to define a second, larger lengthscale, $\xi_{\text{marginal}}$. We also look at correlation in the transverse direction, $C(y)$ at a small, fixed $x$ ($x_{\rm MD}=50$ and $x_{\rm EPM}=40$). We find that $C(y)$ is well modeled by a Gaussian, $C(y)\sim e^{-(y/w)^2/2}$ (results not shown), which allows us to quantify the width $w$.

In Fig.~\ref{fig:plasticlength} (c) and (g), we show that these three lengths exhibit a power-law evolution with $\Delta t$. The trends are similar in both MD and EPM. $\xi_{\text{excite}}$ has the strongest $\Delta t$ dependence: it is scales like $\Delta t^{3/4}$ in the MD and is nearly ballistic in the EPM. $\xi_{\text{marginal}}$ scales diffusively as $\Delta t^{1/2}$ in both the MD and EPM. $w$ has a very weak dependence on $\Delta t$ which indicates that for any given avalanche, the correlations get more an more constrained to lie along the axes of shear as time goes during the course of the avalanche. Finally, in Fig.~\ref{fig:plasticlength} (d) and (h) we show how $\xi_{\text{marginal}}$ and $\xi_{\text{excite}}$ depend on system size for a fixed, small $\Delta t$. In both MD and EPM, $\xi_{\text{excite}}$ has relatively little dependence on system size, while $\xi_{\text{marginal}}$ has a more pronounced system size dependence, scaling approximately like $L^{1/3}$ and $L^{2/3}$ in the MD and EPM respectively.

In this letter, we demonstrate that the spatio-temporal evolution of avalanches in particle-based simulations in the overdamped limit share important similarities and differences with an elasto-plastic model governed by synchronous dynamics. We show that in both models, avalanches are driven by localized clusters of activity, and that the spatial correlation function of activity exhibits two lengthscales that grow differently in time and with system size. 

One lengthscale, $\xi_{excite}$, is mechanically excited where a diffusively propagating stress field triggers nearby sites that generate their own stress fields. In the EPM models, this leads to standard ballistic propagation as seen in other excitable media, while in MD simulations we find the front propagates as $t^{3/4}$. In both cases, our data suggests this scale is independent of system size, as expected. An interesting question for future work is what generates this non-standard exponent in MD simulations. 

A second, longer lengthscale, $\xi_{marginal}$, propagates diffusively and scales with system size. This suggests that it is governed by the weakest spots in the disordered solid, as far-field stress fluctuations anywhere in the system are sufficient to trigger events, and the stress magnitude required to trigger the weakest spot scales as a power law with system size.

Another difference between the two types of simulations is the statistics of their temporal dynamics. In MD simulations, the stress overshoot -- the difference between the triggering stress field and the stress threshold of an excitable site -- governs how fast the system departs a saddle and the waiting times between localized bursts of deformation in an avalanche. These effects generate dynamical exponents that depart significantly with mean-field predictions, and cause duration-size curves to vary with system size. In contrast, our simple EPM has no such mechanism; the dynamical exponent is much closer to mean-field and the duration-size curves do not depend on system size. Recent progressive rate EP models~\cite{aguirre2018critical,ferrero2019criticality} include a proxy for this waiting time, and qualitatively reproduce this physics, but do not quantitatively match the exponents we find in the MD. A nice feature of EPMs is that it is possible to disambiguate the consequences of various choices for the stress propagation, waiting times, and disorder. Future work could focus on adjusting properties of EPMs to improve quantitative agreement with MD. Another important avenue is understanding how these models behave in the presence of inertia~\cite{Salerno-PRL12,Salerno-PRE13,karimi2017inertia}.

We have focused here on MD and EPM simulations for a relatively ductile material, where the stress required for each site to be triggered is relatively small. It will be very interesting to revisit similarities and differences in brittle systems that undergo shear banding instabilities, which provide an even stricter test of models of the spatio-temporal evolution of disordered solids. More broadly, the tools developed here could be used to characterize spatio-temporal dynamics of avalanches in materials with complicated interactions (irregular shapes, friction, realistic molecular potentials) and boundary conditions.

{\bf Acknowledgments} We acknowledge support of the Simons Foundation for the
``Cracking the Glass Problem Collaboration'' Award No.~348126 (D.R) and No.~454947 (A.E. and M.L.M). M.L.M. acknowledges support from NSF-DMR-1951921.\\

% \bibliography{references}

\begin{thebibliography}{49}%
\makeatletter
\providecommand \@ifxundefined [1]{%
 \@ifx{#1\undefined}
}%
\providecommand \@ifnum [1]{%
 \ifnum #1\expandafter \@firstoftwo
 \else \expandafter \@secondoftwo
 \fi
}%
\providecommand \@ifx [1]{%
 \ifx #1\expandafter \@firstoftwo
 \else \expandafter \@secondoftwo
 \fi
}%
\providecommand \natexlab [1]{#1}%
\providecommand \enquote  [1]{``#1''}%
\providecommand \bibnamefont  [1]{#1}%
\providecommand \bibfnamefont [1]{#1}%
\providecommand \citenamefont [1]{#1}%
\providecommand \href@noop [0]{\@secondoftwo}%
\providecommand \href [0]{\begingroup \@sanitize@url \@href}%
\providecommand \@href[1]{\@@startlink{#1}\@@href}%
\providecommand \@@href[1]{\endgroup#1\@@endlink}%
\providecommand \@sanitize@url [0]{\catcode `\\12\catcode `\$12\catcode
  `\&12\catcode `\#12\catcode `\^12\catcode `\_12\catcode `\%12\relax}%
\providecommand \@@startlink[1]{}%
\providecommand \@@endlink[0]{}%
\providecommand \url  [0]{\begingroup\@sanitize@url \@url }%
\providecommand \@url [1]{\endgroup\@href {#1}{\urlprefix }}%
\providecommand \urlprefix  [0]{URL }%
\providecommand \Eprint [0]{\href }%
\providecommand \doibase [0]{https://doi.org/}%
\providecommand \selectlanguage [0]{\@gobble}%
\providecommand \bibinfo  [0]{\@secondoftwo}%
\providecommand \bibfield  [0]{\@secondoftwo}%
\providecommand \translation [1]{[#1]}%
\providecommand \BibitemOpen [0]{}%
\providecommand \bibitemStop [0]{}%
\providecommand \bibitemNoStop [0]{.\EOS\space}%
\providecommand \EOS [0]{\spacefactor3000\relax}%
\providecommand \BibitemShut  [1]{\csname bibitem#1\endcsname}%
\let\auto@bib@innerbib\@empty
%</preamble>
\bibitem [{\citenamefont {Durin}\ and\ \citenamefont
  {Zapperi}(2000)}]{durin2000scaling}%
  \BibitemOpen
  \bibfield  {author} {\bibinfo {author} {\bibfnamefont {G.}~\bibnamefont
  {Durin}}\ and\ \bibinfo {author} {\bibfnamefont {S.}~\bibnamefont
  {Zapperi}},\ }\bibfield  {title} {\bibinfo {title} {Scaling exponents for
  barkhausen avalanches in polycrystalline and amorphous ferromagnets},\ }\href
  {https://doi.org/https://doi.org/10.1103/PhysRevLett.84.4705} {\bibfield
  {journal} {\bibinfo  {journal} {Phys. Rev. Lett.}\ }\textbf {\bibinfo
  {volume} {84}},\ \bibinfo {pages} {4705} (\bibinfo {year}
  {2000})}\BibitemShut {NoStop}%
\bibitem [{\citenamefont {Dalton}\ and\ \citenamefont
  {Corcoran}(2001)}]{dalton2001self}%
  \BibitemOpen
  \bibfield  {author} {\bibinfo {author} {\bibfnamefont {F.}~\bibnamefont
  {Dalton}}\ and\ \bibinfo {author} {\bibfnamefont {D.}~\bibnamefont
  {Corcoran}},\ }\bibfield  {title} {\bibinfo {title} {Self-organized
  criticality in a sheared granular stick-slip system},\ }\href
  {https://doi.org/https://doi.org/10.1103/PhysRevE.63.061312} {\bibfield
  {journal} {\bibinfo  {journal} {Phys. Rev. E}\ }\textbf {\bibinfo {volume}
  {63}},\ \bibinfo {pages} {061312} (\bibinfo {year} {2001})}\BibitemShut
  {NoStop}%
\bibitem [{\citenamefont {M{\aa}l{\o}y}\ \emph {et~al.}(2006)\citenamefont
  {M{\aa}l{\o}y}, \citenamefont {Santucci}, \citenamefont {Schmittbuhl},\ and\
  \citenamefont {Toussaint}}]{maaloy2006local}%
  \BibitemOpen
  \bibfield  {author} {\bibinfo {author} {\bibfnamefont {K.~J.}\ \bibnamefont
  {M{\aa}l{\o}y}}, \bibinfo {author} {\bibfnamefont {S.}~\bibnamefont
  {Santucci}}, \bibinfo {author} {\bibfnamefont {J.}~\bibnamefont
  {Schmittbuhl}},\ and\ \bibinfo {author} {\bibfnamefont {R.}~\bibnamefont
  {Toussaint}},\ }\bibfield  {title} {\bibinfo {title} {Local waiting time
  fluctuations along a randomly pinned crack front},\ }\href
  {https://doi.org/https://doi.org/10.1103/PhysRevLett.96.045501} {\bibfield
  {journal} {\bibinfo  {journal} {Phys. Rev. Lett.}\ }\textbf {\bibinfo
  {volume} {96}},\ \bibinfo {pages} {045501} (\bibinfo {year}
  {2006})}\BibitemShut {NoStop}%
\bibitem [{\citenamefont {Sethna}\ \emph {et~al.}(2001)\citenamefont {Sethna},
  \citenamefont {Dahmen},\ and\ \citenamefont {Myers}}]{sethna2001crackling}%
  \BibitemOpen
  \bibfield  {author} {\bibinfo {author} {\bibfnamefont {J.~P.}\ \bibnamefont
  {Sethna}}, \bibinfo {author} {\bibfnamefont {K.~A.}\ \bibnamefont {Dahmen}},\
  and\ \bibinfo {author} {\bibfnamefont {C.~R.}\ \bibnamefont {Myers}},\
  }\bibfield  {title} {\bibinfo {title} {Crackling noise},\ }\href
  {https://doi.org/https://doi.org/10.1038/35065675} {\bibfield  {journal}
  {\bibinfo  {journal} {Nature}\ }\textbf {\bibinfo {volume} {410}},\ \bibinfo
  {pages} {242} (\bibinfo {year} {2001})}\BibitemShut {NoStop}%
\bibitem [{\citenamefont {Sun}\ \emph {et~al.}(2010)\citenamefont {Sun},
  \citenamefont {Yu}, \citenamefont {Jiao}, \citenamefont {Bai}, \citenamefont
  {Zhao},\ and\ \citenamefont {Wang}}]{sun2010plasticity}%
  \BibitemOpen
  \bibfield  {author} {\bibinfo {author} {\bibfnamefont {B.}~\bibnamefont
  {Sun}}, \bibinfo {author} {\bibfnamefont {H.}~\bibnamefont {Yu}}, \bibinfo
  {author} {\bibfnamefont {W.}~\bibnamefont {Jiao}}, \bibinfo {author}
  {\bibfnamefont {H.}~\bibnamefont {Bai}}, \bibinfo {author} {\bibfnamefont
  {D.}~\bibnamefont {Zhao}},\ and\ \bibinfo {author} {\bibfnamefont
  {W.}~\bibnamefont {Wang}},\ }\bibfield  {title} {\bibinfo {title} {Plasticity
  of ductile metallic glasses: A self-organized critical state},\ }\href
  {https://doi.org/https://doi.org/10.1103/PhysRevLett.105.035501} {\bibfield
  {journal} {\bibinfo  {journal} {Phys. Rev. Lett.}\ }\textbf {\bibinfo
  {volume} {105}},\ \bibinfo {pages} {035501} (\bibinfo {year}
  {2010})}\BibitemShut {NoStop}%
\bibitem [{\citenamefont {Lin}\ \emph {et~al.}(2015)\citenamefont {Lin},
  \citenamefont {Gueudr{\'e}}, \citenamefont {Rosso},\ and\ \citenamefont
  {Wyart}}]{lin2015criticality}%
  \BibitemOpen
  \bibfield  {author} {\bibinfo {author} {\bibfnamefont {J.}~\bibnamefont
  {Lin}}, \bibinfo {author} {\bibfnamefont {T.}~\bibnamefont {Gueudr{\'e}}},
  \bibinfo {author} {\bibfnamefont {A.}~\bibnamefont {Rosso}},\ and\ \bibinfo
  {author} {\bibfnamefont {M.}~\bibnamefont {Wyart}},\ }\bibfield  {title}
  {\bibinfo {title} {Criticality in the approach to failure in amorphous
  solids},\ }\href
  {https://doi.org/https://doi.org/10.1103/PhysRevLett.115.168001} {\bibfield
  {journal} {\bibinfo  {journal} {Phys. Rev. Lett.}\ }\textbf {\bibinfo
  {volume} {115}},\ \bibinfo {pages} {168001} (\bibinfo {year}
  {2015})}\BibitemShut {NoStop}%
\bibitem [{\citenamefont {Budrikis}\ \emph {et~al.}(2017)\citenamefont
  {Budrikis}, \citenamefont {Castellanos}, \citenamefont {Sandfeld},
  \citenamefont {Zaiser},\ and\ \citenamefont {Zapperi}}]{Budrikis-NatComm17}%
  \BibitemOpen
  \bibfield  {author} {\bibinfo {author} {\bibfnamefont {Z.}~\bibnamefont
  {Budrikis}}, \bibinfo {author} {\bibfnamefont {D.~F.}\ \bibnamefont
  {Castellanos}}, \bibinfo {author} {\bibfnamefont {S.}~\bibnamefont
  {Sandfeld}}, \bibinfo {author} {\bibfnamefont {M.}~\bibnamefont {Zaiser}},\
  and\ \bibinfo {author} {\bibfnamefont {S.}~\bibnamefont {Zapperi}},\
  }\bibfield  {title} {\bibinfo {title} {Universal features of amorphous
  plasticity},\ }\href {https://doi.org/https://doi.org/10.1038/ncomms15928}
  {\bibfield  {journal} {\bibinfo  {journal} {Nature Comm.}\ }\textbf {\bibinfo
  {volume} {8}},\ \bibinfo {pages} {15928} (\bibinfo {year}
  {2017})}\BibitemShut {NoStop}%
\bibitem [{\citenamefont {Talamali}\ \emph {et~al.}(2011)\citenamefont
  {Talamali}, \citenamefont {Pet\"aj\"a}, \citenamefont {Vandembroucq},\ and\
  \citenamefont {Roux}}]{Talamali-PRE11}%
  \BibitemOpen
  \bibfield  {author} {\bibinfo {author} {\bibfnamefont {M.}~\bibnamefont
  {Talamali}}, \bibinfo {author} {\bibfnamefont {V.}~\bibnamefont
  {Pet\"aj\"a}}, \bibinfo {author} {\bibfnamefont {D.}~\bibnamefont
  {Vandembroucq}},\ and\ \bibinfo {author} {\bibfnamefont {S.}~\bibnamefont
  {Roux}},\ }\bibfield  {title} {\bibinfo {title} {Avalanches, precursors and
  finite size fluctuations in a mesoscopic model of amorphous plasticity},\
  }\href {https://doi.org/https://doi.org/10.1103/PhysRevE.84.016115}
  {\bibfield  {journal} {\bibinfo  {journal} {Phys. Rev. E}\ }\textbf {\bibinfo
  {volume} {84}},\ \bibinfo {pages} {016115} (\bibinfo {year}
  {2011})}\BibitemShut {NoStop}%
\bibitem [{\citenamefont {Salerno}\ \emph {et~al.}(2012)\citenamefont
  {Salerno}, \citenamefont {Maloney},\ and\ \citenamefont
  {Robbins}}]{Salerno-PRL12}%
  \BibitemOpen
  \bibfield  {author} {\bibinfo {author} {\bibfnamefont {K.~M.}\ \bibnamefont
  {Salerno}}, \bibinfo {author} {\bibfnamefont {C.~E.}\ \bibnamefont
  {Maloney}},\ and\ \bibinfo {author} {\bibfnamefont {M.~O.}\ \bibnamefont
  {Robbins}},\ }\bibfield  {title} {\bibinfo {title} {Avalanches in strained
  amorphous solids: Does inertia destroy critical behavior?},\ }\href
  {https://doi.org/https://doi.org/10.1103/PhysRevLett.109.105703} {\bibfield
  {journal} {\bibinfo  {journal} {Phys. Rev. Lett.}\ }\textbf {\bibinfo
  {volume} {109}},\ \bibinfo {pages} {105703} (\bibinfo {year}
  {2012})}\BibitemShut {NoStop}%
\bibitem [{\citenamefont {Lin}\ \emph {et~al.}(2014{\natexlab{a}})\citenamefont
  {Lin}, \citenamefont {Lerner}, \citenamefont {Rosso},\ and\ \citenamefont
  {Wyart}}]{lin2014scaling}%
  \BibitemOpen
  \bibfield  {author} {\bibinfo {author} {\bibfnamefont {J.}~\bibnamefont
  {Lin}}, \bibinfo {author} {\bibfnamefont {E.}~\bibnamefont {Lerner}},
  \bibinfo {author} {\bibfnamefont {A.}~\bibnamefont {Rosso}},\ and\ \bibinfo
  {author} {\bibfnamefont {M.}~\bibnamefont {Wyart}},\ }\bibfield  {title}
  {\bibinfo {title} {Scaling description of the yielding transition in soft
  amorphous solids at zero temperature},\ }\href
  {https://doi.org/https://doi.org/10.1073/pnas.1406391111} {\bibfield
  {journal} {\bibinfo  {journal} {Proceedings of the National Academy of
  Sciences}\ }\textbf {\bibinfo {volume} {111}},\ \bibinfo {pages} {14382}
  (\bibinfo {year} {2014}{\natexlab{a}})}\BibitemShut {NoStop}%
\bibitem [{\citenamefont {Liu}\ \emph {et~al.}(2016)\citenamefont {Liu},
  \citenamefont {Ferrero}, \citenamefont {Puosi}, \citenamefont {Barrat},\ and\
  \citenamefont {Martens}}]{liu2016driving}%
  \BibitemOpen
  \bibfield  {author} {\bibinfo {author} {\bibfnamefont {C.}~\bibnamefont
  {Liu}}, \bibinfo {author} {\bibfnamefont {E.~E.}\ \bibnamefont {Ferrero}},
  \bibinfo {author} {\bibfnamefont {F.}~\bibnamefont {Puosi}}, \bibinfo
  {author} {\bibfnamefont {J.-L.}\ \bibnamefont {Barrat}},\ and\ \bibinfo
  {author} {\bibfnamefont {K.}~\bibnamefont {Martens}},\ }\bibfield  {title}
  {\bibinfo {title} {Driving rate dependence of avalanche statistics and shapes
  at the yielding transition},\ }\href
  {https://journals.aps.org/prl/abstract/10.1103/PhysRevLett.116.065501}
  {\bibfield  {journal} {\bibinfo  {journal} {Phys. Rev. lett.}\ }\textbf
  {\bibinfo {volume} {116}},\ \bibinfo {pages} {065501} (\bibinfo {year}
  {2016})}\BibitemShut {NoStop}%
\bibitem [{\citenamefont {Clemmer}\ \emph
  {et~al.}(2021{\natexlab{a}})\citenamefont {Clemmer}, \citenamefont
  {Salerno},\ and\ \citenamefont {Robbins}}]{clemmer2021criticalityp1}%
  \BibitemOpen
  \bibfield  {author} {\bibinfo {author} {\bibfnamefont {J.~T.}\ \bibnamefont
  {Clemmer}}, \bibinfo {author} {\bibfnamefont {K.~M.}\ \bibnamefont
  {Salerno}},\ and\ \bibinfo {author} {\bibfnamefont {M.~O.}\ \bibnamefont
  {Robbins}},\ }\bibfield  {title} {\bibinfo {title} {Criticality in sheared,
  disordered solids. i. rate effects in stress and diffusion},\ }\href
  {https://doi.org/https://doi.org/10.1103/PhysRevE.103.042605} {\bibfield
  {journal} {\bibinfo  {journal} {Phys. Rev. E}\ }\textbf {\bibinfo {volume}
  {103}},\ \bibinfo {pages} {042605} (\bibinfo {year}
  {2021}{\natexlab{a}})}\BibitemShut {NoStop}%
\bibitem [{\citenamefont {Clemmer}\ \emph
  {et~al.}(2021{\natexlab{b}})\citenamefont {Clemmer}, \citenamefont
  {Salerno},\ and\ \citenamefont {Robbins}}]{clemmer2021criticalityp2}%
  \BibitemOpen
  \bibfield  {author} {\bibinfo {author} {\bibfnamefont {J.~T.}\ \bibnamefont
  {Clemmer}}, \bibinfo {author} {\bibfnamefont {K.~M.}\ \bibnamefont
  {Salerno}},\ and\ \bibinfo {author} {\bibfnamefont {M.~O.}\ \bibnamefont
  {Robbins}},\ }\bibfield  {title} {\bibinfo {title} {Criticality in sheared,
  disordered solids. ii. correlations in avalanche dynamics},\ }\href
  {https://doi.org/https://doi.org/10.1103/PhysRevE.103.042606} {\bibfield
  {journal} {\bibinfo  {journal} {Phys. Rev. E}\ }\textbf {\bibinfo {volume}
  {103}},\ \bibinfo {pages} {042606} (\bibinfo {year}
  {2021}{\natexlab{b}})}\BibitemShut {NoStop}%
\bibitem [{\citenamefont {Maloney}\ and\ \citenamefont
  {Robbins}(2009)}]{maloney2009anisotropic}%
  \BibitemOpen
  \bibfield  {author} {\bibinfo {author} {\bibfnamefont {C.}~\bibnamefont
  {Maloney}}\ and\ \bibinfo {author} {\bibfnamefont {M.}~\bibnamefont
  {Robbins}},\ }\bibfield  {title} {\bibinfo {title} {Anisotropic power law
  strain correlations in sheared amorphous 2d solids},\ }\href
  {https://doi.org/https://doi.org/10.1103/PhysRevLett.102.225502} {\bibfield
  {journal} {\bibinfo  {journal} {Phys. Rev. Lett.}\ }\textbf {\bibinfo
  {volume} {102}},\ \bibinfo {pages} {225502} (\bibinfo {year}
  {2009})}\BibitemShut {NoStop}%
\bibitem [{\citenamefont {Talamali}\ \emph {et~al.}(2012)\citenamefont
  {Talamali}, \citenamefont {Pet{\"a}j{\"a}}, \citenamefont {Vandembroucq},\
  and\ \citenamefont {Roux}}]{TALAMALI2012275}%
  \BibitemOpen
  \bibfield  {author} {\bibinfo {author} {\bibfnamefont {M.}~\bibnamefont
  {Talamali}}, \bibinfo {author} {\bibfnamefont {V.}~\bibnamefont
  {Pet{\"a}j{\"a}}}, \bibinfo {author} {\bibfnamefont {D.}~\bibnamefont
  {Vandembroucq}},\ and\ \bibinfo {author} {\bibfnamefont {S.}~\bibnamefont
  {Roux}},\ }\bibfield  {title} {\bibinfo {title} {Strain localization and
  anisotropic correlations in a mesoscopic model of amorphous plasticity},\
  }\href {https://doi.org/https://doi.org/10.1016/j.crme.2012.02.010}
  {\bibfield  {journal} {\bibinfo  {journal} {Comptes Rendus M{\'e}canique}\
  }\textbf {\bibinfo {volume} {340}},\ \bibinfo {pages} {275} (\bibinfo {year}
  {2012})},\ \bibinfo {note} {recent Advances in Micromechanics of
  Materials}\BibitemShut {NoStop}%
\bibitem [{\citenamefont {Chattoraj}\ and\ \citenamefont
  {Lema\^{i}tre}(2013)}]{chattoraj2013elastic}%
  \BibitemOpen
  \bibfield  {author} {\bibinfo {author} {\bibfnamefont {J.}~\bibnamefont
  {Chattoraj}}\ and\ \bibinfo {author} {\bibfnamefont {A.}~\bibnamefont
  {Lema\^{i}tre}},\ }\bibfield  {title} {\bibinfo {title} {Elastic signature of
  flow events in supercooled liquids under shear},\ }\href
  {https://doi.org/https://doi.org/10.1103/PhysRevLett.111.066001} {\bibfield
  {journal} {\bibinfo  {journal} {Phys. Rev. Lett.}\ }\textbf {\bibinfo
  {volume} {111}},\ \bibinfo {pages} {066001} (\bibinfo {year}
  {2013})}\BibitemShut {NoStop}%
\bibitem [{\citenamefont {Nicolas}\ \emph {et~al.}(2014)\citenamefont
  {Nicolas}, \citenamefont {Rottler},\ and\ \citenamefont
  {Barrat}}]{nicolas2014spatiotemporal}%
  \BibitemOpen
  \bibfield  {author} {\bibinfo {author} {\bibfnamefont {A.}~\bibnamefont
  {Nicolas}}, \bibinfo {author} {\bibfnamefont {J.}~\bibnamefont {Rottler}},\
  and\ \bibinfo {author} {\bibfnamefont {J.-L.}\ \bibnamefont {Barrat}},\
  }\bibfield  {title} {\bibinfo {title} {Spatiotemporal correlations between
  plastic events in the shear flow of athermal amorphous solids},\ }\href
  {https://doi.org/https://doi.org/10.1140/epje/i2014-14050-1} {\bibfield
  {journal} {\bibinfo  {journal} {Eur. Phys. J. E}\ }\textbf {\bibinfo {volume}
  {37}},\ \bibinfo {pages} {1} (\bibinfo {year} {2014})}\BibitemShut {NoStop}%
\bibitem [{\citenamefont {Puosi}\ \emph {et~al.}(2016)\citenamefont {Puosi},
  \citenamefont {Rottler},\ and\ \citenamefont {Barrat}}]{puosi2016plastic}%
  \BibitemOpen
  \bibfield  {author} {\bibinfo {author} {\bibfnamefont {F.}~\bibnamefont
  {Puosi}}, \bibinfo {author} {\bibfnamefont {J.}~\bibnamefont {Rottler}},\
  and\ \bibinfo {author} {\bibfnamefont {J.-L.}\ \bibnamefont {Barrat}},\
  }\bibfield  {title} {\bibinfo {title} {Plastic response and correlations in
  athermally sheared amorphous solids},\ }\href
  {https://doi.org/https://doi.org/10.1103/PhysRevE.94.032604} {\bibfield
  {journal} {\bibinfo  {journal} {Phys. Rev. E}\ }\textbf {\bibinfo {volume}
  {94}},\ \bibinfo {pages} {032604} (\bibinfo {year} {2016})}\BibitemShut
  {NoStop}%
\bibitem [{\citenamefont {Antonaglia}\ \emph {et~al.}(2014)\citenamefont
  {Antonaglia}, \citenamefont {Wright}, \citenamefont {Gu}, \citenamefont
  {Byer}, \citenamefont {Hufnagel}, \citenamefont {LeBlanc}, \citenamefont
  {Uhl},\ and\ \citenamefont {Dahmen}}]{Dahmen-PRL14}%
  \BibitemOpen
  \bibfield  {author} {\bibinfo {author} {\bibfnamefont {J.}~\bibnamefont
  {Antonaglia}}, \bibinfo {author} {\bibfnamefont {W.~J.}\ \bibnamefont
  {Wright}}, \bibinfo {author} {\bibfnamefont {X.}~\bibnamefont {Gu}}, \bibinfo
  {author} {\bibfnamefont {R.~R.}\ \bibnamefont {Byer}}, \bibinfo {author}
  {\bibfnamefont {T.~C.}\ \bibnamefont {Hufnagel}}, \bibinfo {author}
  {\bibfnamefont {M.}~\bibnamefont {LeBlanc}}, \bibinfo {author} {\bibfnamefont
  {J.~T.}\ \bibnamefont {Uhl}},\ and\ \bibinfo {author} {\bibfnamefont {K.~A.}\
  \bibnamefont {Dahmen}},\ }\bibfield  {title} {\bibinfo {title} {Bulk metallic
  glasses deform via slip avalanches},\ }\href
  {https://doi.org/https://doi.org/10.1103/PhysRevLett.112.155501} {\bibfield
  {journal} {\bibinfo  {journal} {Phys. Rev. Lett.}\ }\textbf {\bibinfo
  {volume} {112}},\ \bibinfo {pages} {155501} (\bibinfo {year}
  {2014})}\BibitemShut {NoStop}%
\bibitem [{\citenamefont {Nicolas}\ \emph {et~al.}(2018)\citenamefont
  {Nicolas}, \citenamefont {Ferrero}, \citenamefont {Martens},\ and\
  \citenamefont {Barrat}}]{nicolas2018deformation}%
  \BibitemOpen
  \bibfield  {author} {\bibinfo {author} {\bibfnamefont {A.}~\bibnamefont
  {Nicolas}}, \bibinfo {author} {\bibfnamefont {E.~E.}\ \bibnamefont
  {Ferrero}}, \bibinfo {author} {\bibfnamefont {K.}~\bibnamefont {Martens}},\
  and\ \bibinfo {author} {\bibfnamefont {J.-L.}\ \bibnamefont {Barrat}},\
  }\bibfield  {title} {\bibinfo {title} {Deformation and flow of amorphous
  solids: Insights from elastoplastic models},\ }\href
  {https://doi.org/https://doi.org/10.1103/RevModPhys.90.045006} {\bibfield
  {journal} {\bibinfo  {journal} {RMP}\ }\textbf {\bibinfo {volume} {90}},\
  \bibinfo {pages} {045006} (\bibinfo {year} {2018})}\BibitemShut {NoStop}%
\bibitem [{\citenamefont {Baret}\ \emph {et~al.}(2002)\citenamefont {Baret},
  \citenamefont {Vandembroucq},\ and\ \citenamefont {Roux}}]{baret-PRL02}%
  \BibitemOpen
  \bibfield  {author} {\bibinfo {author} {\bibfnamefont {J.-C.}\ \bibnamefont
  {Baret}}, \bibinfo {author} {\bibfnamefont {D.}~\bibnamefont
  {Vandembroucq}},\ and\ \bibinfo {author} {\bibfnamefont {S.}~\bibnamefont
  {Roux}},\ }\bibfield  {title} {\bibinfo {title} {An extremal model of
  amorphous plasticity},\ }\href
  {https://doi.org/https://doi.org/10.1103/PhysRevLett.89.195506} {\bibfield
  {journal} {\bibinfo  {journal} {Phys. Rev. Lett.}\ }\textbf {\bibinfo
  {volume} {89}},\ \bibinfo {pages} {195506} (\bibinfo {year}
  {2002})}\BibitemShut {NoStop}%
\bibitem [{\citenamefont {Budrikis}\ and\ \citenamefont
  {Zapperi}(2013)}]{Budrikis-PRE13}%
  \BibitemOpen
  \bibfield  {author} {\bibinfo {author} {\bibfnamefont {Z.}~\bibnamefont
  {Budrikis}}\ and\ \bibinfo {author} {\bibfnamefont {S.}~\bibnamefont
  {Zapperi}},\ }\bibfield  {title} {\bibinfo {title} {Avalanche localization
  and crossover scaling in amorphous plasticity},\ }\href
  {https://doi.org/https://doi.org/10.1103/PhysRevE.88.062403} {\bibfield
  {journal} {\bibinfo  {journal} {Phys. Rev. E}\ }\textbf {\bibinfo {volume}
  {88}},\ \bibinfo {pages} {062403} (\bibinfo {year} {2013})}\BibitemShut
  {NoStop}%
\bibitem [{\citenamefont {Tyukodi}\ \emph {et~al.}(2018)\citenamefont
  {Tyukodi}, \citenamefont {Vandembroucq},\ and\ \citenamefont
  {Maloney}}]{Tyukodi-PRL18}%
  \BibitemOpen
  \bibfield  {author} {\bibinfo {author} {\bibfnamefont {B.}~\bibnamefont
  {Tyukodi}}, \bibinfo {author} {\bibfnamefont {D.}~\bibnamefont
  {Vandembroucq}},\ and\ \bibinfo {author} {\bibfnamefont {C.~E.}\ \bibnamefont
  {Maloney}},\ }\bibfield  {title} {\bibinfo {title} {Diffusion in mesoscopic
  lattice models of amorphous plasticity},\ }\href
  {https://doi.org/https://doi.org/10.1103/PhysRevLett.121.145501} {\bibfield
  {journal} {\bibinfo  {journal} {Phys. Rev. Lett.}\ }\textbf {\bibinfo
  {volume} {121}},\ \bibinfo {pages} {145501} (\bibinfo {year}
  {2018})}\BibitemShut {NoStop}%
\bibitem [{\citenamefont {Puosi}\ \emph {et~al.}(2014)\citenamefont {Puosi},
  \citenamefont {Rottler},\ and\ \citenamefont {Barrat}}]{puosi2014time}%
  \BibitemOpen
  \bibfield  {author} {\bibinfo {author} {\bibfnamefont {F.}~\bibnamefont
  {Puosi}}, \bibinfo {author} {\bibfnamefont {J.}~\bibnamefont {Rottler}},\
  and\ \bibinfo {author} {\bibfnamefont {J.-L.}\ \bibnamefont {Barrat}},\
  }\bibfield  {title} {\bibinfo {title} {Time-dependent elastic response to a
  local shear transformation in amorphous solids},\ }\href
  {https://doi.org/https://doi.org/10.1103/PhysRevE.89.042302} {\bibfield
  {journal} {\bibinfo  {journal} {Phys. Rev. E}\ }\textbf {\bibinfo {volume}
  {89}},\ \bibinfo {pages} {042302} (\bibinfo {year} {2014})}\BibitemShut
  {NoStop}%
\bibitem [{\citenamefont {Salerno}\ and\ \citenamefont
  {Robbins}(2013)}]{Salerno-PRE13}%
  \BibitemOpen
  \bibfield  {author} {\bibinfo {author} {\bibfnamefont {K.~M.}\ \bibnamefont
  {Salerno}}\ and\ \bibinfo {author} {\bibfnamefont {M.~O.}\ \bibnamefont
  {Robbins}},\ }\bibfield  {title} {\bibinfo {title} {Effect of inertia on
  sheared disordered solids: critical scaling of avalanches in two and three
  dimensions},\ }\href
  {https://doi.org/https://doi.org/10.1103/PhysRevE.88.062206} {\bibfield
  {journal} {\bibinfo  {journal} {Phys. Rev. E}\ }\textbf {\bibinfo {volume}
  {88}},\ \bibinfo {pages} {062206} (\bibinfo {year} {2013})}\BibitemShut
  {NoStop}%
\bibitem [{\citenamefont {Tyukodi}\ \emph {et~al.}(2019)\citenamefont
  {Tyukodi}, \citenamefont {Vandembroucq},\ and\ \citenamefont
  {Maloney}}]{Tyukodi-PRE19}%
  \BibitemOpen
  \bibfield  {author} {\bibinfo {author} {\bibfnamefont {B.}~\bibnamefont
  {Tyukodi}}, \bibinfo {author} {\bibfnamefont {D.}~\bibnamefont
  {Vandembroucq}},\ and\ \bibinfo {author} {\bibfnamefont {C.~E.}\ \bibnamefont
  {Maloney}},\ }\bibfield  {title} {\bibinfo {title} {Avalanches, thresholds,
  and diffusion in mesoscale amorphous plasticity},\ }\href
  {https://doi.org/https://doi.org/10.1103/PhysRevE.100.043003} {\bibfield
  {journal} {\bibinfo  {journal} {Phys. Rev. E}\ }\textbf {\bibinfo {volume}
  {100}},\ \bibinfo {pages} {043003} (\bibinfo {year} {2019})}\BibitemShut
  {NoStop}%
\bibitem [{\citenamefont {Idema}\ and\ \citenamefont
  {Liu}(2014)}]{idema2014mechanical}%
  \BibitemOpen
  \bibfield  {author} {\bibinfo {author} {\bibfnamefont {T.}~\bibnamefont
  {Idema}}\ and\ \bibinfo {author} {\bibfnamefont {A.~J.}\ \bibnamefont
  {Liu}},\ }\bibfield  {title} {\bibinfo {title} {Mechanical signaling via
  nonlinear wavefront propagation in a mechanically excitable medium},\ }\href
  {https://doi.org/https://doi.org/10.1103/PhysRevE.89.062709} {\bibfield
  {journal} {\bibinfo  {journal} {Phys. Rev. E}\ }\textbf {\bibinfo {volume}
  {89}},\ \bibinfo {pages} {062709} (\bibinfo {year} {2014})}\BibitemShut
  {NoStop}%
\bibitem [{\citenamefont {Sollich}\ \emph {et~al.}(1997)\citenamefont
  {Sollich}, \citenamefont {Lequeux}, \citenamefont {H\'ebraud},\ and\
  \citenamefont {Cates}}]{Sollich-PRL97}%
  \BibitemOpen
  \bibfield  {author} {\bibinfo {author} {\bibfnamefont {P.}~\bibnamefont
  {Sollich}}, \bibinfo {author} {\bibfnamefont {F.}~\bibnamefont {Lequeux}},
  \bibinfo {author} {\bibfnamefont {P.}~\bibnamefont {H\'ebraud}},\ and\
  \bibinfo {author} {\bibfnamefont {M.~E.}\ \bibnamefont {Cates}},\ }\bibfield
  {title} {\bibinfo {title} {Rheology of soft glassy materials},\ }\href
  {https://doi.org/https://doi.org/10.1103/PhysRevLett.78.2020} {\bibfield
  {journal} {\bibinfo  {journal} {Phys. Rev. Lett.}\ }\textbf {\bibinfo
  {volume} {78}},\ \bibinfo {pages} {2020} (\bibinfo {year}
  {1997})}\BibitemShut {NoStop}%
\bibitem [{\citenamefont {Lema\^{i}tre}\ and\ \citenamefont
  {Caroli}(2007)}]{Lemaitre-Caroli-PRE07}%
  \BibitemOpen
  \bibfield  {author} {\bibinfo {author} {\bibfnamefont {A.}~\bibnamefont
  {Lema\^{i}tre}}\ and\ \bibinfo {author} {\bibfnamefont {C.}~\bibnamefont
  {Caroli}},\ }\href
  {https://doi.org/https://doi.org/10.1103/PhysRevE.76.036104} {\bibfield
  {journal} {\bibinfo  {journal} {Phys. Rev. E}\ }\textbf {\bibinfo {volume}
  {76}},\ \bibinfo {pages} {036104} (\bibinfo {year} {2007})}\BibitemShut
  {NoStop}%
\bibitem [{\citenamefont {Agoritsas}\ \emph {et~al.}(2015)\citenamefont
  {Agoritsas}, \citenamefont {Bertin}, \citenamefont {Martens},\ and\
  \citenamefont {Barrat}}]{Agoritsas-EPJE15}%
  \BibitemOpen
  \bibfield  {author} {\bibinfo {author} {\bibfnamefont {E.}~\bibnamefont
  {Agoritsas}}, \bibinfo {author} {\bibfnamefont {E.}~\bibnamefont {Bertin}},
  \bibinfo {author} {\bibfnamefont {K.}~\bibnamefont {Martens}},\ and\ \bibinfo
  {author} {\bibfnamefont {J.-L.}\ \bibnamefont {Barrat}},\ }\bibfield  {title}
  {\bibinfo {title} {On the relevance of disorder in athermal amorphous
  materials under shear},\ }\href
  {https://doi.org/https://doi.org/10.1140/epje/i2015-15071-x} {\bibfield
  {journal} {\bibinfo  {journal} {Eur. Phys. J. E}\ }\textbf {\bibinfo {volume}
  {38}} (\bibinfo {year} {2015})}\BibitemShut {NoStop}%
\bibitem [{\citenamefont {Lin}\ and\ \citenamefont
  {Wyart}(2016)}]{Wyart-PRX16}%
  \BibitemOpen
  \bibfield  {author} {\bibinfo {author} {\bibfnamefont {J.}~\bibnamefont
  {Lin}}\ and\ \bibinfo {author} {\bibfnamefont {M.}~\bibnamefont {Wyart}},\
  }\bibfield  {title} {\bibinfo {title} {Mean-field description of plastic flow
  in amorphous solids},\ }\href
  {https://doi.org/https://doi.org/10.1103/PhysRevX.6.011005} {\bibfield
  {journal} {\bibinfo  {journal} {Phys. Rev. X}\ }\textbf {\bibinfo {volume}
  {6}},\ \bibinfo {pages} {011005} (\bibinfo {year} {2016})}\BibitemShut
  {NoStop}%
\bibitem [{\citenamefont {Ferrero}\ and\ \citenamefont
  {Jagla}(2019)}]{ferrero2019criticality}%
  \BibitemOpen
  \bibfield  {author} {\bibinfo {author} {\bibfnamefont {E.~E.}\ \bibnamefont
  {Ferrero}}\ and\ \bibinfo {author} {\bibfnamefont {E.~A.}\ \bibnamefont
  {Jagla}},\ }\bibfield  {title} {\bibinfo {title} {Criticality in
  elastoplastic models of amorphous solids with stress-dependent yielding
  rates},\ }\href {https://doi.org/10.1039/C9SM01073D} {\bibfield  {journal}
  {\bibinfo  {journal} {Soft matter}\ }\textbf {\bibinfo {volume} {15}},\
  \bibinfo {pages} {9041} (\bibinfo {year} {2019})}\BibitemShut {NoStop}%
\bibitem [{\citenamefont {Lerner}(2019)}]{lerner2019mechanical}%
  \BibitemOpen
  \bibfield  {author} {\bibinfo {author} {\bibfnamefont {E.}~\bibnamefont
  {Lerner}},\ }\bibfield  {title} {\bibinfo {title} {Mechanical properties of
  simple computer glasses},\ }\href
  {https://doi.org/https://doi.org/10.1016/j.jnoncrysol.2019.119570} {\bibfield
   {journal} {\bibinfo  {journal} {J. Non-Cryst. Solids}\ }\textbf {\bibinfo
  {volume} {522}},\ \bibinfo {pages} {119570} (\bibinfo {year}
  {2019})}\BibitemShut {NoStop}%
\bibitem [{\citenamefont {Maloney}\ and\ \citenamefont
  {Lemaitre}(2006)}]{maloney2006amorphous}%
  \BibitemOpen
  \bibfield  {author} {\bibinfo {author} {\bibfnamefont {C.~E.}\ \bibnamefont
  {Maloney}}\ and\ \bibinfo {author} {\bibfnamefont {A.}~\bibnamefont
  {Lemaitre}},\ }\bibfield  {title} {\bibinfo {title} {Amorphous systems in
  athermal, quasistatic shear},\ }\href
  {https://doi.org/https://doi.org/10.1103/PhysRevE.74.016118} {\bibfield
  {journal} {\bibinfo  {journal} {Phys. Rev. E}\ }\textbf {\bibinfo {volume}
  {74}},\ \bibinfo {pages} {016118} (\bibinfo {year} {2006})}\BibitemShut
  {NoStop}%
\bibitem [{\citenamefont {Stanifer}\ and\ \citenamefont
  {Manning}(2021)}]{stanifer2021avalanche}%
  \BibitemOpen
  \bibfield  {author} {\bibinfo {author} {\bibfnamefont {E.}~\bibnamefont
  {Stanifer}}\ and\ \bibinfo {author} {\bibfnamefont {M.~L.}\ \bibnamefont
  {Manning}},\ }\bibfield  {title} {\bibinfo {title} {Avalanche dynamics in
  sheared athermal particle packings occurs via localized bursts predicted by
  unstable linear response},\ }\href {https://arxiv.org/abs/2110.02803}
  {\bibfield  {journal} {\bibinfo  {journal} {arXiv preprint arXiv:2110.02803}\
  } (\bibinfo {year} {2021})}\BibitemShut {NoStop}%
\bibitem [{SM()}]{SM}%
  \BibitemOpen
  \href@noop {} {}\bibinfo {note} {See Supplemental Material at
  http://xxx.yyy.zzz for details.}\BibitemShut {Stop}%
\bibitem [{\citenamefont {Khirallah}\ \emph {et~al.}(2021)\citenamefont
  {Khirallah}, \citenamefont {Tyukodi}, \citenamefont {Vandembroucq},\ and\
  \citenamefont {Maloney}}]{Khirallah2021yielding}%
  \BibitemOpen
  \bibfield  {author} {\bibinfo {author} {\bibfnamefont {K.}~\bibnamefont
  {Khirallah}}, \bibinfo {author} {\bibfnamefont {B.}~\bibnamefont {Tyukodi}},
  \bibinfo {author} {\bibfnamefont {D.}~\bibnamefont {Vandembroucq}},\ and\
  \bibinfo {author} {\bibfnamefont {C.~E.}\ \bibnamefont {Maloney}},\
  }\bibfield  {title} {\bibinfo {title} {Yielding in an integer automaton model
  for amorphous solids under cyclic shear},\ }\href
  {https://link.aps.org/doi/10.1103/PhysRevLett.126.218005} {\bibfield
  {journal} {\bibinfo  {journal} {Phys. Rev. Lett.}\ }\textbf {\bibinfo
  {volume} {126}},\ \bibinfo {pages} {218005} (\bibinfo {year}
  {2021})}\BibitemShut {NoStop}%
\bibitem [{\citenamefont {Laurson}\ \emph {et~al.}(2010)\citenamefont
  {Laurson}, \citenamefont {Santucci},\ and\ \citenamefont
  {Zapperi}}]{laurson2010avalanches}%
  \BibitemOpen
  \bibfield  {author} {\bibinfo {author} {\bibfnamefont {L.}~\bibnamefont
  {Laurson}}, \bibinfo {author} {\bibfnamefont {S.}~\bibnamefont {Santucci}},\
  and\ \bibinfo {author} {\bibfnamefont {S.}~\bibnamefont {Zapperi}},\
  }\bibfield  {title} {\bibinfo {title} {Avalanches and clusters in planar
  crack front propagation},\ }\href
  {https://doi.org/https://doi.org/10.1103/PhysRevE.81.046116} {\bibfield
  {journal} {\bibinfo  {journal} {Phys. Rev. E}\ }\textbf {\bibinfo {volume}
  {81}},\ \bibinfo {pages} {046116} (\bibinfo {year} {2010})}\BibitemShut
  {NoStop}%
\bibitem [{\citenamefont {Le~Priol}\ \emph {et~al.}(2021)\citenamefont
  {Le~Priol}, \citenamefont {Le~Doussal},\ and\ \citenamefont
  {Rosso}}]{le2021spatial}%
  \BibitemOpen
  \bibfield  {author} {\bibinfo {author} {\bibfnamefont {C.}~\bibnamefont
  {Le~Priol}}, \bibinfo {author} {\bibfnamefont {P.}~\bibnamefont
  {Le~Doussal}},\ and\ \bibinfo {author} {\bibfnamefont {A.}~\bibnamefont
  {Rosso}},\ }\bibfield  {title} {\bibinfo {title} {Spatial clustering of
  depinning avalanches in presence of long-range interactions},\ }\href
  {https://doi.org/https://doi.org/10.1103/PhysRevLett.126.025702} {\bibfield
  {journal} {\bibinfo  {journal} {Phys. Rev. Lett.}\ }\textbf {\bibinfo
  {volume} {126}},\ \bibinfo {pages} {025702} (\bibinfo {year}
  {2021})}\BibitemShut {NoStop}%
\bibitem [{\citenamefont {Strogatz}(2018)}]{strogatz2018nonlinear}%
  \BibitemOpen
  \bibfield  {author} {\bibinfo {author} {\bibfnamefont {S.~H.}\ \bibnamefont
  {Strogatz}},\ }\href {https://doi.org/https://doi.org/10.1201/9780429399640}
  {\emph {\bibinfo {title} {Nonlinear dynamics and chaos with student solutions
  manual: With applications to physics, biology, chemistry, and engineering}}}\
  (\bibinfo  {publisher} {CRC press},\ \bibinfo {year} {2018})\BibitemShut
  {NoStop}%
\bibitem [{\citenamefont {Aguirre}\ and\ \citenamefont
  {Jagla}(2018)}]{aguirre2018critical}%
  \BibitemOpen
  \bibfield  {author} {\bibinfo {author} {\bibfnamefont {I.~F.}\ \bibnamefont
  {Aguirre}}\ and\ \bibinfo {author} {\bibfnamefont {E.~A.}\ \bibnamefont
  {Jagla}},\ }\bibfield  {title} {\bibinfo {title} {Critical exponents of the
  yielding transition of amorphous solids},\ }\href
  {https://doi.org/https://doi.org/10.1103/PhysRevE.98.013002} {\bibfield
  {journal} {\bibinfo  {journal} {Phys. Rev. E}\ }\textbf {\bibinfo {volume}
  {98}},\ \bibinfo {pages} {013002} (\bibinfo {year} {2018})}\BibitemShut
  {NoStop}%
\bibitem [{\citenamefont {Karmakar}\ \emph {et~al.}(2010)\citenamefont
  {Karmakar}, \citenamefont {Lerner},\ and\ \citenamefont
  {Procaccia}}]{karmakar2010statistical}%
  \BibitemOpen
  \bibfield  {author} {\bibinfo {author} {\bibfnamefont {S.}~\bibnamefont
  {Karmakar}}, \bibinfo {author} {\bibfnamefont {E.}~\bibnamefont {Lerner}},\
  and\ \bibinfo {author} {\bibfnamefont {I.}~\bibnamefont {Procaccia}},\
  }\bibfield  {title} {\bibinfo {title} {Statistical physics of the yielding
  transition in amorphous solids},\ }\href
  {https://doi.org/https://doi.org/10.1103/PhysRevE.82.055103} {\bibfield
  {journal} {\bibinfo  {journal} {Phys. Rev. E}\ }\textbf {\bibinfo {volume}
  {82}},\ \bibinfo {pages} {055103} (\bibinfo {year} {2010})}\BibitemShut
  {NoStop}%
\bibitem [{\citenamefont {Lin}\ \emph {et~al.}(2014{\natexlab{b}})\citenamefont
  {Lin}, \citenamefont {Saade}, \citenamefont {Lerner}, \citenamefont {Rosso},\
  and\ \citenamefont {Wyart}}]{lin2014density}%
  \BibitemOpen
  \bibfield  {author} {\bibinfo {author} {\bibfnamefont {J.}~\bibnamefont
  {Lin}}, \bibinfo {author} {\bibfnamefont {A.}~\bibnamefont {Saade}}, \bibinfo
  {author} {\bibfnamefont {E.}~\bibnamefont {Lerner}}, \bibinfo {author}
  {\bibfnamefont {A.}~\bibnamefont {Rosso}},\ and\ \bibinfo {author}
  {\bibfnamefont {M.}~\bibnamefont {Wyart}},\ }\bibfield  {title} {\bibinfo
  {title} {On the density of shear transformations in amorphous solids},\
  }\href {https://doi.org/10.1209/0295-5075/105/26003} {\bibfield  {journal}
  {\bibinfo  {journal} {EPL}\ }\textbf {\bibinfo {volume} {105}},\ \bibinfo
  {pages} {26003} (\bibinfo {year} {2014}{\natexlab{b}})}\BibitemShut {NoStop}%
\bibitem [{\citenamefont {Ruscher}\ and\ \citenamefont
  {Rottler}(2020)}]{ruscher2020residual}%
  \BibitemOpen
  \bibfield  {author} {\bibinfo {author} {\bibfnamefont {C.}~\bibnamefont
  {Ruscher}}\ and\ \bibinfo {author} {\bibfnamefont {J.}~\bibnamefont
  {Rottler}},\ }\bibfield  {title} {\bibinfo {title} {Residual stress
  distributions in amorphous solids from atomistic simulations},\ }\href
  {https://doi.org/DOI https://doi.org/10.1039/D0SM01155J} {\bibfield
  {journal} {\bibinfo  {journal} {Soft Matter}\ }\textbf {\bibinfo {volume}
  {16}},\ \bibinfo {pages} {8940} (\bibinfo {year} {2020})}\BibitemShut
  {NoStop}%
\bibitem [{\citenamefont {Laurson}\ \emph {et~al.}(2013)\citenamefont
  {Laurson}, \citenamefont {Illa}, \citenamefont {Santucci}, \citenamefont
  {Tallakstad}, \citenamefont {M{\aa}l{\o}y},\ and\ \citenamefont
  {Alava}}]{laurson2013evolution}%
  \BibitemOpen
  \bibfield  {author} {\bibinfo {author} {\bibfnamefont {L.}~\bibnamefont
  {Laurson}}, \bibinfo {author} {\bibfnamefont {X.}~\bibnamefont {Illa}},
  \bibinfo {author} {\bibfnamefont {S.}~\bibnamefont {Santucci}}, \bibinfo
  {author} {\bibfnamefont {K.~T.}\ \bibnamefont {Tallakstad}}, \bibinfo
  {author} {\bibfnamefont {K.~J.}\ \bibnamefont {M{\aa}l{\o}y}},\ and\ \bibinfo
  {author} {\bibfnamefont {M.~J.}\ \bibnamefont {Alava}},\ }\bibfield  {title}
  {\bibinfo {title} {Evolution of the average avalanche shape with the
  universality class},\ }\href
  {https://doi.org/https://doi.org/10.1038/ncomms3927} {\bibfield  {journal}
  {\bibinfo  {journal} {Nat. Commun.}\ }\textbf {\bibinfo {volume} {4}},\
  \bibinfo {pages} {1} (\bibinfo {year} {2013})}\BibitemShut {NoStop}%
\bibitem [{\citenamefont {Baldassarri}\ \emph {et~al.}(2019)\citenamefont
  {Baldassarri}, \citenamefont {Annunziata}, \citenamefont {Gnoli},
  \citenamefont {Pontuale},\ and\ \citenamefont
  {Petri}}]{baldassarri2019breakdown}%
  \BibitemOpen
  \bibfield  {author} {\bibinfo {author} {\bibfnamefont {A.}~\bibnamefont
  {Baldassarri}}, \bibinfo {author} {\bibfnamefont {M.}~\bibnamefont
  {Annunziata}}, \bibinfo {author} {\bibfnamefont {A.}~\bibnamefont {Gnoli}},
  \bibinfo {author} {\bibfnamefont {G.}~\bibnamefont {Pontuale}},\ and\
  \bibinfo {author} {\bibfnamefont {A.}~\bibnamefont {Petri}},\ }\bibfield
  {title} {\bibinfo {title} {Breakdown of scaling and friction weakening in
  intermittent granular flow},\ }\href
  {https://doi.org/https://doi.org/10.1038/s41598-019-53178-2} {\bibfield
  {journal} {\bibinfo  {journal} {Scientific reports}\ }\textbf {\bibinfo
  {volume} {9}},\ \bibinfo {pages} {1} (\bibinfo {year} {2019})}\BibitemShut
  {NoStop}%
\bibitem [{not()}]{notesymmetry}%
  \BibitemOpen
  \href@noop {} {}\bibinfo {note} {Note that small deviations from an ideal $x
  \leftrightarrow y$ symmetry stem from the use of Lees-Edwards periodic
  boundary conditions.}\BibitemShut {Stop}%
\bibitem [{\citenamefont {Karimi}\ \emph {et~al.}(2017)\citenamefont {Karimi},
  \citenamefont {Ferrero},\ and\ \citenamefont {Barrat}}]{karimi2017inertia}%
  \BibitemOpen
  \bibfield  {author} {\bibinfo {author} {\bibfnamefont {K.}~\bibnamefont
  {Karimi}}, \bibinfo {author} {\bibfnamefont {E.~E.}\ \bibnamefont
  {Ferrero}},\ and\ \bibinfo {author} {\bibfnamefont {J.-L.}\ \bibnamefont
  {Barrat}},\ }\bibfield  {title} {\bibinfo {title} {Inertia and universality
  of avalanche statistics: The case of slowly deformed amorphous solids},\
  }\href {https://doi.org/https://doi.org/10.1103/PhysRevE.95.013003}
  {\bibfield  {journal} {\bibinfo  {journal} {Phys. Rev. E}\ }\textbf {\bibinfo
  {volume} {95}},\ \bibinfo {pages} {013003} (\bibinfo {year}
  {2017})}\BibitemShut {NoStop}%
\bibitem [{\citenamefont {Falk}\ and\ \citenamefont
  {Langer}(1998)}]{falk1998dynamics}%
  \BibitemOpen
  \bibfield  {author} {\bibinfo {author} {\bibfnamefont {M.~L.}\ \bibnamefont
  {Falk}}\ and\ \bibinfo {author} {\bibfnamefont {J.~S.}\ \bibnamefont
  {Langer}},\ }\bibfield  {title} {\bibinfo {title} {Dynamics of viscoplastic
  deformation in amorphous solids},\ }\href
  {https://doi.org/https://doi.org/10.1103/PhysRevE.57.7192} {\bibfield
  {journal} {\bibinfo  {journal} {Phys. Rev. E}\ }\textbf {\bibinfo {volume}
  {57}},\ \bibinfo {pages} {7192} (\bibinfo {year} {1998})}\BibitemShut
  {NoStop}%
\end{thebibliography}


\onecolumngrid
\cleardoublepage
\begin{center}
	\textbf{\large Supplementary material: ``Mechanical excitation and marginal triggering during avalanches in sheared amorphous solids''}
\end{center}

%%%%%%%%%%%%%%%%%%%%%%%%%%%%%%%%%%%%%%%%%%%%%%%%%%%%%%%%%%%%%%%%%%%%%%%%%%%%%%%%%
%%%%%%%%%%%%%%%%%%%%%% these lines of code handle the concatenation %%%%%%%%%%%%%
%%%%%%%%%%%%%%%%%%%%%%%%%%%%%%%%%%%%%%%%%%%%%%%%%%%%%%%%%%%%%%%%%%%%%%%%%%%%%%%%%
\setcounter{equation}{0}
\setcounter{figure}{0}
\setcounter{section}{0}
\setcounter{table}{0}
\setcounter{page}{1}
\makeatletter
\renewcommand{\theequation}{S\arabic{equation}}
\renewcommand{\thefigure}{S\arabic{figure}}
\renewcommand{\thesubsection}{S-\arabic{subsection}}
\renewcommand{\thesection}{S-\arabic{section}}
\renewcommand*{\thepage}{S\arabic{page}}
\renewcommand{\bibnumfmt}[1]{[S#1]}
\renewcommand{\citenumfont}[1]{S#1}
%%%%%%%%%%%%%%%%%%%%%%%%%%%%%%%%%%%%%%%%%%%%%%%%%%%%%%%%%%%%%%%%%%%%%%%%%%%%%%%%%
%%%%%%%%%%%%%%%%%%%%%% these lines of code handle the concatenation %%%%%%%%%%%%%
%%%%%%%%%%%%%%%%%%%%%%%%%%%%%%%%%%%%%%%%%%%%%%%%%%%%%%%%%%%%%%%%%%%%%%%%%%%%%%%%%

%%%%%%%%%%%%%%%%%%%%%%%%%%%%%%%%%%%%%%%%%%%%%%%%%%%%%%%%%%%%%%%%%%%%%%%%%%%%%%%%%
%%%%%%%%%%%%%%%%%%%%%% these lines of code handle the concatenation %%%%%%%%%%%%%
%%%%%%%%%%%%%%%%%%%%%%%%%%%%%%%%%%%%%%%%%%%%%%%%%%%%%%%%%%%%%%%%%%%%%%%%%%%%%%%%%
\setcounter{equation}{0}
\setcounter{figure}{0}
\setcounter{section}{0}
\setcounter{table}{0}
\setcounter{page}{1}
\makeatletter
\renewcommand{\theequation}{S\arabic{equation}}
\renewcommand{\thefigure}{S\arabic{figure}}
\renewcommand{\thesection}{S-\Roman{section}}
\renewcommand*{\thepage}{S\arabic{page}}
\renewcommand{\bibnumfmt}[1]{[S#1]}
\renewcommand{\citenumfont}[1]{S#1}
%%%%%%%%%%%%%%%%%%%%%%%%%%%%%%%%%%%%%%%%%%%%%%%%%%%%%%%%%%%%%%%%%%%%%%%%%%%%%%%%%
%%%%%%%%%%%%%%%%%%%%%% these lines of code handle the concatenation %%%%%%%%%%%%%
%%%%%%%%%%%%%%%%%%%%%%%%%%%%%%%%%%%%%%%%%%%%%%%%%%%%%%%%%%%%%%%%%%%%%%%%%%%%%%%%%

\section*{1. Description of molecular dynamics and synchronous elastoplastic model}
\label{sm:EPM}

Our MD glass former consists of a standard two dimensional 50:50 binary mixture of `large' and `small' particles of equal mass interacting via a short-range non-additive inverse-power-law potential (see details in Ref.~\cite{lerner2019mechanical}). We have considered different system sizes $L=80,160,320$, and $640$, respectively. Glasses are athermal and quasistatically sheared (AQS) up to 50\% of strain (with strain step $\Delta\gamma=10^{-4}$), and data reported here correspond to strains ranging from 20 to 50 \%. Variations in strain step are discussed below. At the onset of each instability, we trigger the avalanche by affinely deforming the system and subsequently let the system relax via gradient descent dynamics according to $\dot{\rv}=-D\nabla U$, with coordinate positions $\rv$ and potential energy $U$. During the minimization process, we monitor the plastic deformation (i.e. the incremental non-affine displacement) in the standard way by computing the $D^2_{\rm min}$ field~\cite{falk1998dynamics} between snapshots separated by $2\tau_{MD}$, where the microscopic time is defined as $\tau_{MD}=\sigma^2/D$ with $\sigma$ a typical particle diameter and $D$ the bare translational diffusion coefficient. We have collected 20k and 1.2k avalanche dynamics for the smallest ($L=80$) and largest ($L=640$) system, respectively.

Here we provide a synopsis of the elasto-plastic model used in this paper, which was developed in previous work~\cite{Khirallah2021yielding}. A 2D plane is discretized into sites. Each site has an elastic strain $\epsilon_e$ and a plastic strain $\epsilon_p$, with the total strain defined as the sum of the two: $\epsilon_t = \epsilon_e + \epsilon_p$. The stress $\sigma$, is proportional to $\epsilon_e: \sigma = 2\mu \epsilon_e$, where $\sigma$ is defined in terms of the Cartesian components of the Cauchy stress: $\sigma = (\sigma_{xx} - \sigma_{yy})/2$ and, similarly, $\epsilon_e = (\epsilon_e^{xx} - \epsilon_e^{yy})/2$. We set $\mu = 1$ throughout this study, so $\sigma$ and $\epsilon_e$ are numerically equal. When the magnitude of the stress at a site exceeds the plastic threshold, set to 1 here, we increment the plastic strain at the site by in the same direction as $\sigma$ ( +2 for $\sigma \ge 1$, -2 for $\sigma \le -1$), and update the stress at the remaining sites according to the rules of linear elasticity analogous to Eshelby’s classical solution for a plastic inclusion in an elastic matrix. The increment value 2 is chosen to ensure a single-valued strain energy function. The total number of avalanches collected is 36.8K; obtained from shearing 50 different systems (L=256).

\section*{2. Persistent homology methodology}
\label{sm:persistent}

\begin{figure*}[h!]
  \includegraphics[width = 0.95\columnwidth]{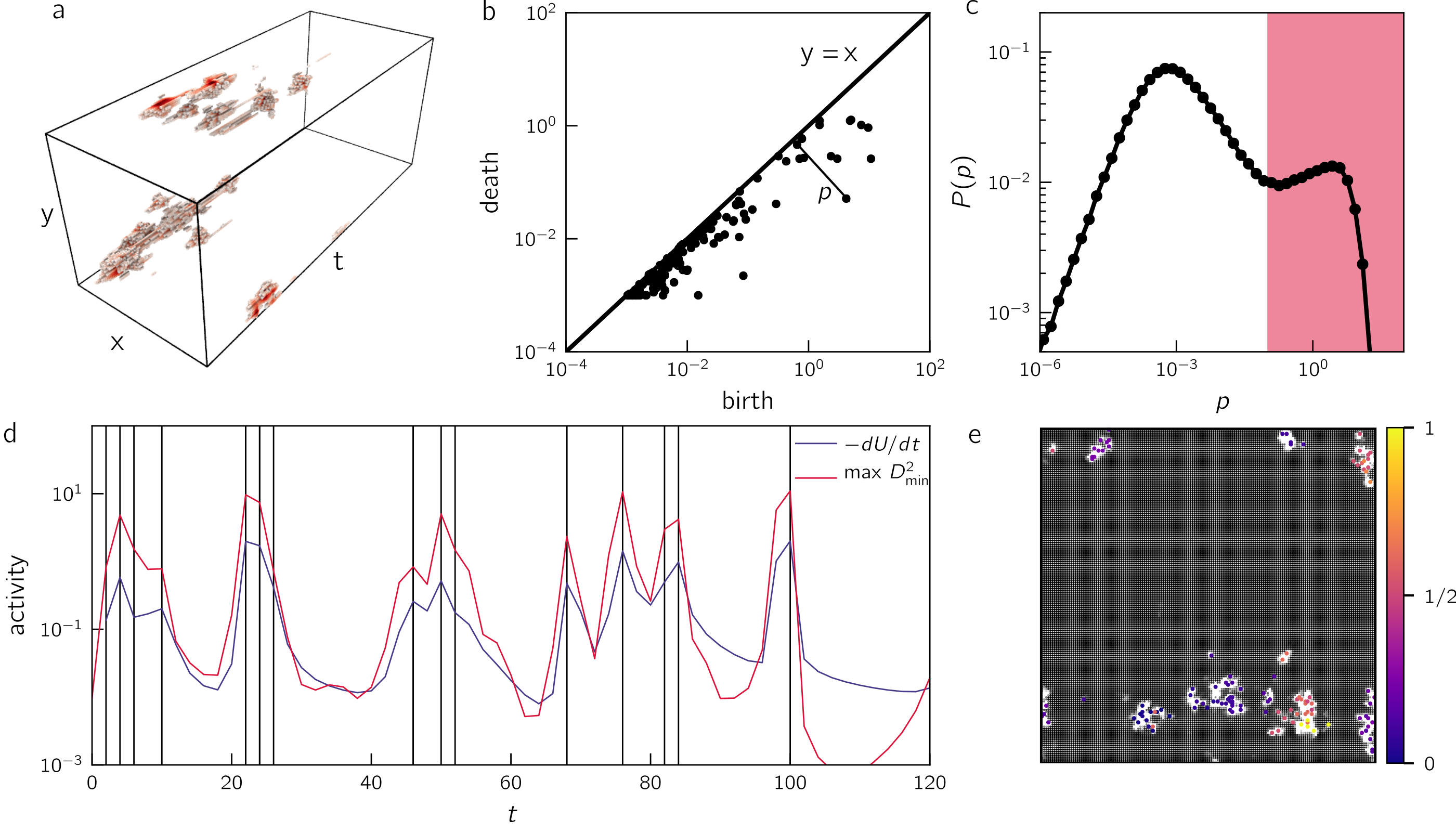} %scale=1.0
    \caption{\textbf{Persistent homology method.} (a) Space-time map of the plastic activity monitored by the $D^2_{\rm min}$ field for a typical avalanche in a glass with $N=160\times160$. (b) Birth-death phase diagram of individual local maxima of our $D^2_{\rm min}$ grid. The persistent $p$ of a given event is given as the distance to the line $y(\rm death)=x(\rm birth)$. (c) Probability distribution of the persistent averaged over our ensemble of avalanches. The red area indicates persistent events with $p>0.1$. (d) Superposition of the actual dissipation signal and birth time of individual events with $p>0.1$. (e) Reconstruction of the spatio-temporal evolution of an avalanche as a sequence of individual events compared to the actual cumulative $D^2_{min}$ plastic field (gray colormap). Colorbar indicates the normalized birth time $t/T$ of each event with $T$ being the avalanche duration.}
  \label{figsm:persistent}
\end{figure*}

In this section, we describe a persistent homology method, similar to that recently proposed by Stanifer and Manning~\cite{stanifer2021avalanche}, employed to break up an avalanche into a time sequence of individual events (as shown in Figure 1 of the main text). In the main text, we refer to the location of each individual event as a "site". During each avalanche, we record the incremental $D^2_{min}$ plastic field which generates a representation of the activity on a grid of space-time regions, see Fig.~\ref{figsm:persistent}(a). The spatial grid size is set to 2 particles diameters and the temporal grid size is set to $2$ MD time units. The $D^2_{min}$ of a cell corresponds to the sum over all $D^2_{min}$ particle values within the cell. Our goal is to agnostically break-up high local maxima of this space-time field. To this end, we employ a persistent homology framework. Here, we first sort all $D^2_{min}$ values from the largest to the lowest up to $10^{-3}$ (from which below no actual plastic rearrangements occur). We then prune this sorted list and create a new cluster for each local maxima that do not belong to an existing cluster, where the clustering distance criterion is set to the first neighbors on the space-time grid. When a new cluster is born we record its $D^2_{min}$ value as its birth. When two existing clusters merge we record their $D^2_{min}$ value as their death. We can then place each cluster in a birth-death diagram (Fig.~\ref{figsm:persistent}(b)) and compute their persistent $p$ as the distance of the cluster to the line $y(\rm death)=x(\rm birth)$. Events that corresponds to short spatime-time fluctuation will be close to this line. When looking at the probability distribution of persistence $P(p)$ average over many avalanches (Fig.~\ref{figsm:persistent}(c)), we observe a bimodal distribution that indicates two different populations of individual events. Selecting events that have a large persistence $p>0.1$ and superimposing their birth time on the avalanche activity time profile (energy dissipation $-dU/dt$ or maximum of the incremental $D^2_{min}$ field), we find that persistent cluster match perfectly the time location of large dissipation. For each persistent cluster, we have access to its spatial position in addition to its birth time. This allows us to construct the same spatio-temporal map of plastic activity as in our EPM, with individual events occurring at "sites" corresponding to these persistent peaks. In Fig.~\ref{figsm:persistent}(d), we show for one avalanche the superposition of the total cumulative plastic field (grayscale colormap) and each individual event colored according to their normalized birth time $t/T$, where $T$ is the total duration of the avalanche.

%%%%%%%%%%%%%%%%%%%%%%%%%%%%%%%% Figure 1 %%%%%%%%%%%%%%%%%%%%%%%%%%%%%
%%%%%%%%%%%%%%%%%%%%%%%%%%%%%%%%%%%%%%%%%%%%%%%%%%%%%%%%%%%%%%%%%%%%%%

\section*{3. Effect of AQS strain step on the avalanche statistics and dynamics}
\label{sm:strainstep}

\begin{figure*}[h!]
  \includegraphics[width = 0.95\columnwidth]{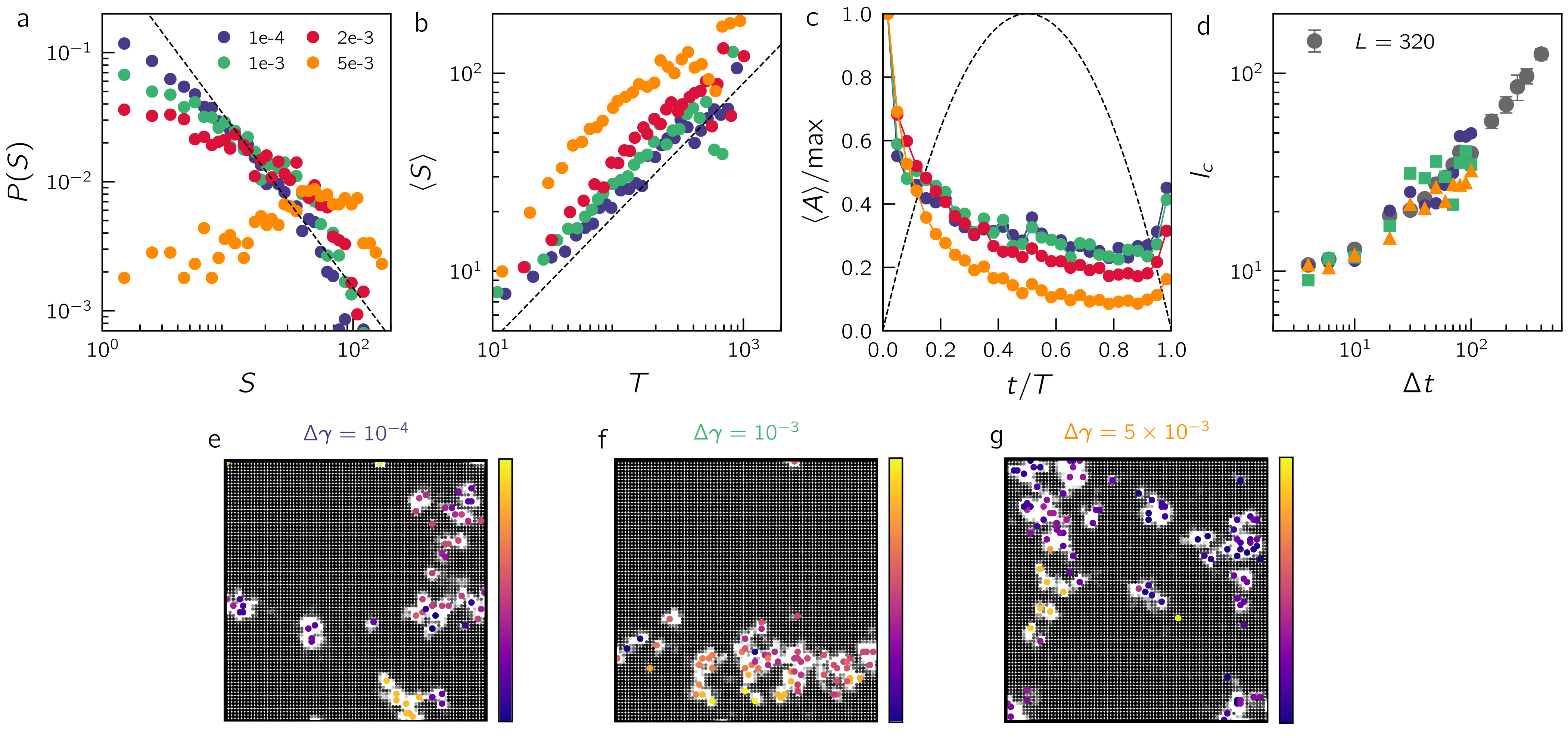} %scale=1.0
    \caption{\textbf{Effect of AQS strain step on the avalanche statistics and dynamics in MD.} (a) Energy drop statistics $P(S)$ for various AQS strain steps $\Delta \gamma$ in the small system $N=160\times 160$. The dashed line indicates power law $P(S)\sim S^{-\tau}$, with $\tau=1.3$. (c) Relation between avlanche size and avalanche duration. The dashed line indicates the relation $\langle S \rangle \sim T^{-\delta}$, with $\delta=0.65$. (d) Plastic length $l_c=\xi_{\text{excite}}$ as a function of the delay time $\Delta t$ extracted from the incremental correlation function (see main text for definition). Different colors correspond to the same strain step values as in (a) and gray points correspond to the data set of a large system with $L=320$. (e), (f), and (g) show typical spatio-temporal evolution of avalanche with $S\simeq100$ for different strain step $\Delta\gamma$. }
  \label{figsm:strainstep}
\end{figure*}

The results presented in the main text are exclusively for AQS dynamics with a strain step $\Delta\gamma=10^{-4}$, which is lower than the average strain distance to the next instability $\langle \Delta \gamma_{\rm min} \rangle=8\times 10^{-4}$. In Fig.\ref{figsm:strainstep}, we show how a large strain step ($\Delta\gamma>\langle \Delta \gamma_{\rm min}\rangle$) can affect the energy drop statistics and avalanche dynamics. In Fig.~\ref{figsm:strainstep}(a), we show the energy drop statistics $P(S)$ for various $\Delta\gamma$. For small $\Delta\gamma$, we found a $S$ range which is consistent with $P(S)\sim S^{-\tau}$, where $\tau=1.3$ such as found in previous studies~\cite{clemmer2021criticalityp1,liu2016driving}. In contrast, as $\Delta\gamma$ increases we observe a depletion of small avalanches in favor of large ones. Here, the effect of a large strain steps is similar as applying a finite strain rate where distinct avalanches, in the true AQS limit, are being glued together. Looking at the avalanche-duration relation $\langle S \rangle \sim T$ shown in Fig.~\ref{figsm:strainstep}(b), we observe that avalanches of the same average size have a shorter duration. This trend is linked to marginal sites that are being pushed further above their threshold, and therefore being activated on a shorter time scale. In Fig.~\ref{figsm:strainstep}(c), we present the normalized average activation profile $\langle A \rangle=\langle -dU/dt\rangle$ for avalanches of size $40<S<100$. We find a significant depletion of activity for $t/T>10\%$ for $\Delta \gamma=5\times 10^{-3}$. In Fig.~\ref{figsm:strainstep}(e), (f), and (g), we show the spatio-temporal evolution of a typical avalanche (with $S\simeq100$) for different strain steps. For $\Delta\gamma=10^{-4}$, we always find that the avalanche dynamics is triggered by a single event. In contrast, for $\Delta \gamma=5\times 10^{-3}$ (Fig.~\ref{figsm:strainstep}(g)), we find many spatially uncorrelated plastic events at the beginning of the avalanche. Nonetheless, investigating the incremental plastic spatio-temporal propagation and plotting the plastic length $l_c=\xi_{\text{excite}}$ as a function of the delay time $\Delta t$, we do not observe a significant influence of the imposed strain step. Finally, comparing the plastic propagation in the smaller system ($L=160$) with the larger one ($L=320$) we do not observe significant finite size effects, see Fig.~\ref{figsm:strainstep}(d). Similar results have been found in our EPM model by driving the system with a finite strain step instead of the true AQS limit for which the strain increment is always equal to the next lowest threshold in the system.

%%%%%%%%%%%%%%%%%%%%%%%%%%%%%%%% Figure 2 %%%%%%%%%%%%%%%%%%%%%%%%%%%%%
%%%%%%%%%%%%%%%%%%%%%%%%%%%%%%%%%%%%%%%%%%%%%%%%%%%%%%%%%%%%%%%%%%%%%%

\section*{4. Static and dynamical exponents}
\label{sm:collapse}

\begin{figure*}[h!]
  \includegraphics[width =1\columnwidth]{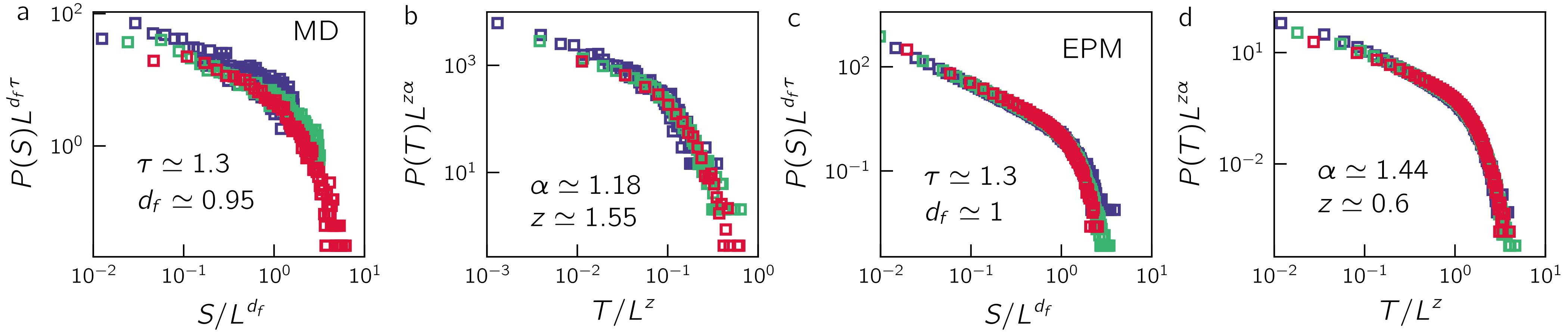} %scale=1.0
    \caption{\textbf{Static and dynamical exponents for MD and EPM.} (a) System size collapse of the MD avalanche size distributions. (b) Same as (a) but for the MD avalanche duration distributions. (c) and (d) show the same results as in (a) and (b) but for our EPM. }
  \label{figsm:exponents}
\end{figure*}

In this section, we extract estimates for the static and dynamical exponents associated with the distribution of avalanche size and duration. We find that our MD avalanche size and duration distributions for different system sizes can be well collapsed using exponents from Ref.~\cite{clemmer2021criticalityp2} , see Fig.~\ref{figsm:exponents}(a) and Fig.~\ref{figsm:exponents}(b). Namely the avalanche statistic follows $P(S)\sim S^{-\tau}$ with $\tau\simeq1.3$ and cutoff $S_c\sim L^{d_f}$, with static fractal dimension $d_f\simeq0.95$ and the duration statistic follows $P(T)\sim T^{-\alpha}$ with $\alpha\simeq1.18$ and cutoff $T_c\sim L^{z}$, with dynamical fractal dimension $z\simeq1.55$. Our EPM shows similar static exponents $\tau\simeq1.3$ and $d_f\simeq1$, but different dynamical exponent $\alpha\simeq1.44$ and $z\simeq0.6$, see Fig.~\ref{figsm:exponents}(c) and Fig.~\ref{figsm:exponents}(d). The latter is consistent with previous estimations using uniform rate and instantaneous stress propagation, see Ref.~\cite{liu2016driving}.

%%%%%%%%%%%%%%%%%%%%%%%%%%%%%%%% Figure 3 %%%%%%%%%%%%%%%%%%%%%%%%%%%%%
%%%%%%%%%%%%%%%%%%%%%%%%%%%%%%%%%%%%%%%%%%%%%%%%%%%%%%%%%%%%%%%%%%%%%%

\section*{5. Typical STZ duration}
\label{sm:stztime}

\begin{figure*}[h!]
  \includegraphics[width = 0.60\columnwidth]{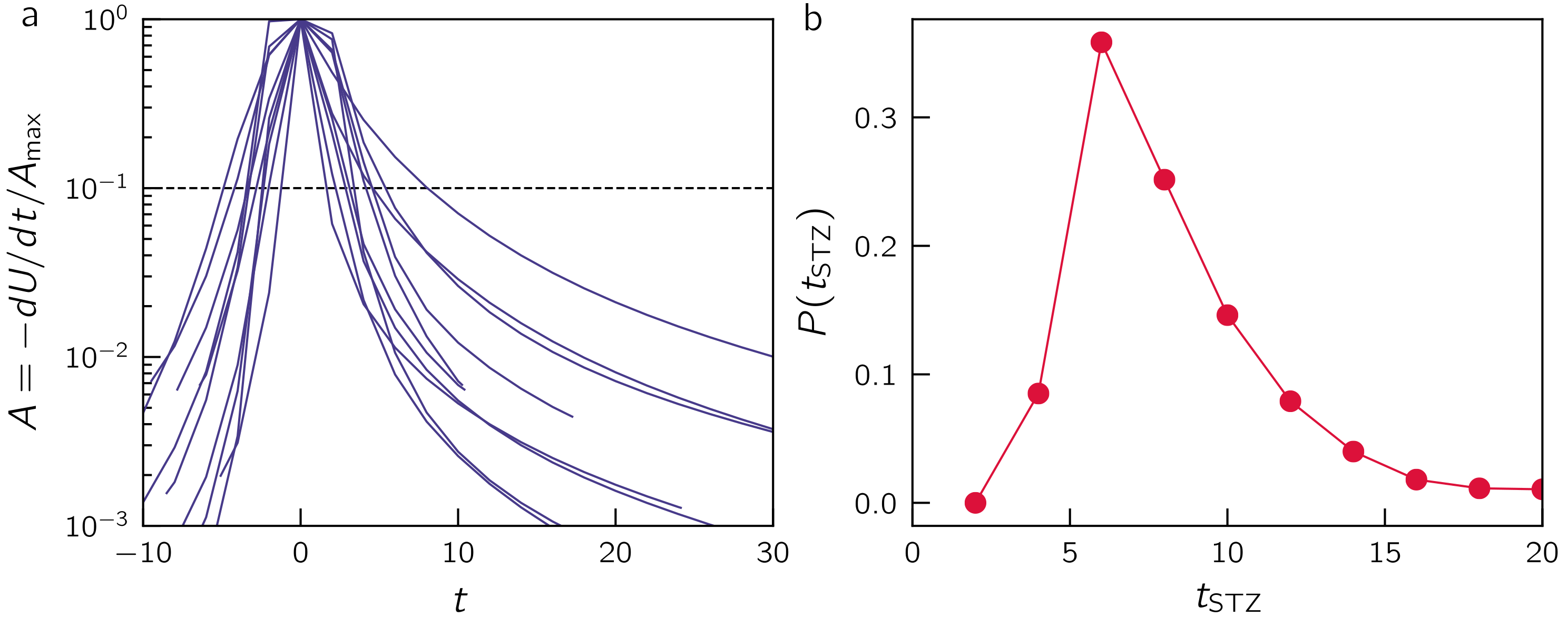} %scale=1.0
    \caption{\textbf{Activity profile of STZ singlets.} (a) Normalized activity profile of several avalanches composed of a single STZ. (b) Distribution of the duration of singlets.}
  \label{figsm:stztime}
\end{figure*}

In contrast to our mesoscale model, there is no analogue to a sweep time in particle-based simulations. Here, we propose to extract a typical duration $t_{\rm STZ}$ associated with an individual STZ. To do so, we select avalanches that are composed of one single peak of activity, as shown in Fig.~\ref{figsm:stztime}(a). We compute the duration of an event as the time for which the activity signal $A(t)$ drops by an order of magnitude from its max $A_{\rm max}$. In Fig.~\ref{figsm:stztime}(b), we show the distribution of duration $P(t_{\rm STZ})$. We estimate a typical STZ rearranges over $6$ units of Brownian time.

%%%%%%%%%%%%%%%%%%%%%%%%%%%%%%%% Figure 4 %%%%%%%%%%%%%%%%%%%%%%%%%%%%%
%%%%%%%%%%%%%%%%%%%%%%%%%%%%%%%%%%%%%%%%%%%%%%%%%%%%%%%%%%%%%%%%%%%%%%

\section*{6. Time translational invariance and size effects}
\label{sm:timeinvariance}

\begin{figure*}[h!]
  \includegraphics[width = 0.9\columnwidth]{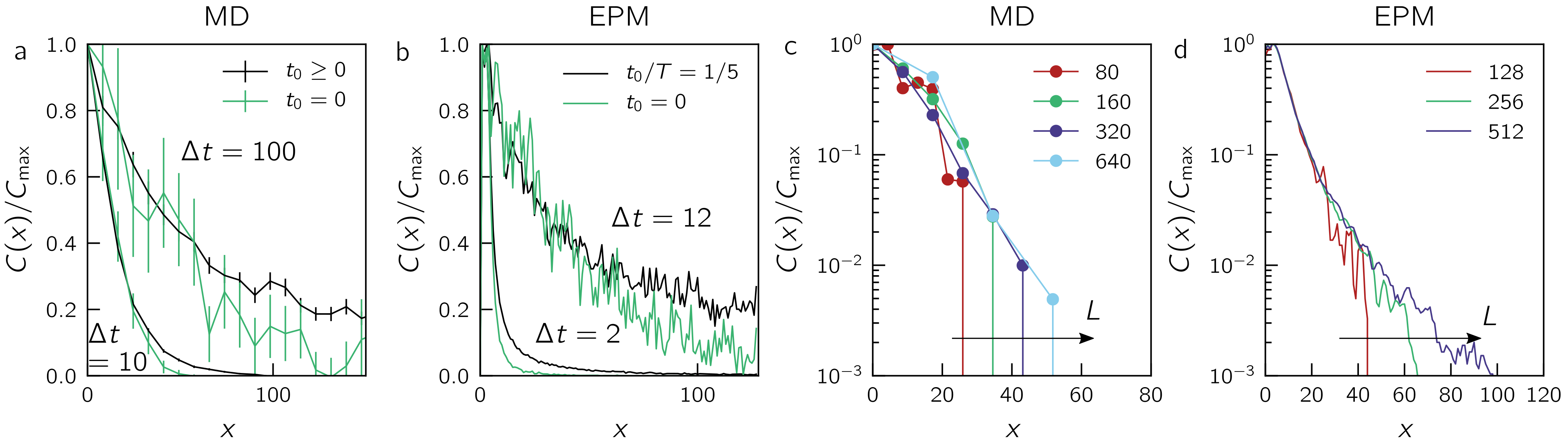} %scale=1.0
    \caption{\textbf{Time translational invariance and size effects in MD and EPM.} (a) Comparison of the MD incremental correlation $C=\langle \overline{\Delta f(r_0,t_0)\Delta f(r_0+\vec{r},t_0+\Delta t)} \rangle$ for $t_0=0$ (green) and  $t_0\ge0$ (black). (b) Same data as in (a) for our EPM but with $t_0=0$ (green) and  $t_0/T=1/5$ (black). (c) MD correlation function at $\Delta t =6$ for different system sizes with box length $L$. (d) Same data as in (c) but for our EPM at $\Delta t =3$}
  \label{figsm:invariance}
\end{figure*}

For both MD and our EPM, we have checked whether or not time translational invariance can be applied to our incremental correlation function, i.e $C(t_0=0)=C(t_0>0)$ where $t_0$ is the reference time of the incremental plastic field that we correlate with subsequent deformation at $t_0+\Delta t$. For both MD (Fig.~\ref{figsm:invariance}(a)) and our EPM (Fig.~\ref{figsm:invariance}(b)), we find that time translational invariance hold relatively well in the early exponential decay. We find a slightly larger likelihood to find more plasticity further in space for $t_0>0$. This trend is explained by the progressive accumulation of stress redistribution across the system, which makes it more likely for marginal sites to yield away from the original source. 

In addition, we show for MD (Fig.~\ref{figsm:invariance}(c)) and our EPM (Fig.~\ref{figsm:invariance}(d)) the influence of the system size for a fixed delay time $\Delta t$. We clearly find for both MD and EPM that the exponential decay does not strongly depend on the system size. In contrast, the negative crossing does increase with the box size $L$, as discussed in more details in the main text.

%%%%%%%%%%%%%%%%%%%%%%%%%%%%%%%% Figure 5 %%%%%%%%%%%%%%%%%%%%%%%%%%%%%
%%%%%%%%%%%%%%%%%%%%%%%%%%%%%%%%%%%%%%%%%%%%%%%%%%%%%%%%%%%%%%%%%%%%%%

\section*{7. Finite size study of $\xi_{\rm excite}$ and $\xi_{\rm marginal}$}
\label{sm:sizeeffect}

In the main text, we only provide the change of $\xi_{\rm excite}$ and $\xi_{\rm marginal}$ as a function of the delay time $\Delta t$ for a fixed system size ($L=320$ for MD and $L=256$ for EPM). Here, we provide additional data for different system sizes. We find that $\xi_{\rm excite}$ does not vary much with $L$. In contrast, $\xi_{\rm marginal}$ grows as the system size increases.

%%%%%%%%%%%%%%%%%%%%%%%%%%%%%%%% Figure 6 %%%%%%%%%%%%%%%%%%%%%%%%%%%%%
\begin{figure*}[h!]
  \includegraphics[width = 1\columnwidth]{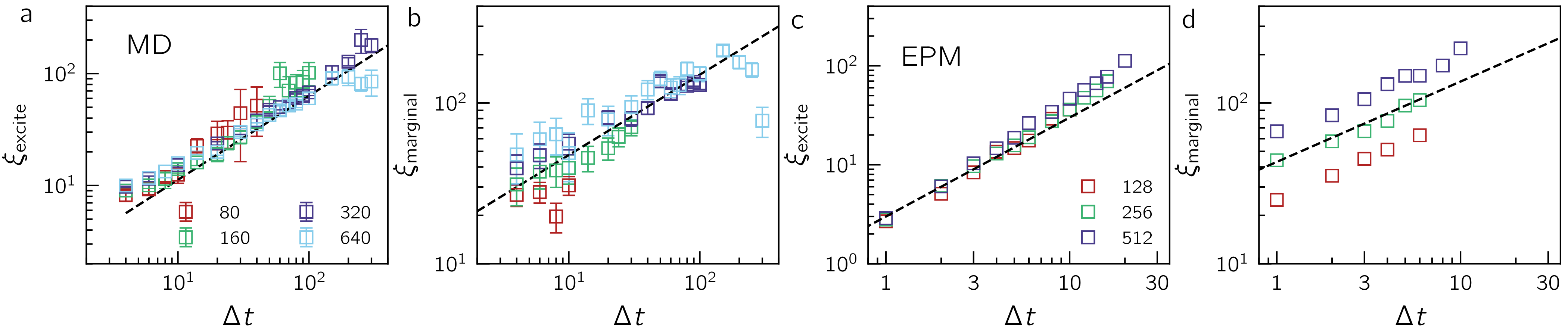} %scale=1.0
    \caption{\textbf{Finite size study of $\xi_{\rm excite}$ and $\xi_{\rm marginal}$.} (a) MD data for $\xi_{\rm excite}$ plotted as a function of the delay time $\Delta t$ for different system size. The dashed line indicates the superdiffusive scaling $t^{3/4}$. (b)$\xi_{\rm marginal}$ as a function of $\Delta t$. The dashed line indicates the diffusive scaling $t^{1/2}$. (c) and (d) are the same as (a) and (b) but for our EPM model. The dashed lines in (c) and (d) indicate $t$ and $t^{1/2}$, respectively.}
  \label{figsm:sizeeffect}
\end{figure*}
%%%%%%%%%%%%%%%%%%%%%%%%%%%%%%%%%%%%%%%%%%%%%%%%%%%%%%%%%%%%%%%%%%%%%%

\end{document}